\documentclass{jfm}

\usepackage{graphicx}
\usepackage{amsmath}
\usepackage{amssymb}
\usepackage{mathrsfs}
\usepackage{bm}
\usepackage{natbib}
\usepackage{hyperref}
\hypersetup{
	colorlinks = true,
	allcolors = blue,
	citecolor  = black,
}
\usepackage{empheq}
\usepackage{xcolor}

\usepackage[width=\linewidth, justification=justified]{caption}

\newcommand{\sub}[1]{{#1}}
\renewcommand\d{\mathrm{d}}
\newcommand\dd{\,\mathrm{d}}
\renewcommand\i{\mathrm{i}}
\newcommand\e{\mathrm{e}}
\newcommand\ohn{\mbox{\textit{Oh}}}
\newcommand\thn{\mbox{\textit{Th}}}
\newcommand{\kb}{k_\sub{B}}
\newcommand{\HL}{H_\sub{LE}}
\newcommand{\HF}{H_\sub{fluc}}
\newcommand{\df}{\rmDelta_\sub{f}}
\newcommand{\db}{\rmDelta_\sub{b}}

\newcommand{\RomanNumeralCaps}[1]
\linenumbers

\shorttitle{Fluctuation-driven dynamics of liquid nano-threads.}
\shortauthor{Z. Zhang, C. Zhao, T. Si}

\title{Fluctuation-driven dynamics of liquid nano-threads with external hydrodynamic perturbations}

\author{
	Zhao Zhang\aff{}%
	,
	Chengxi Zhao\aff{}
	\corresp{\email{zhaochengxi@ustc.edu.cn}}
	\and
	Ting Si\aff{}
}

\affiliation{
	\aff{}Department of Modern Mechanics, 
		University of Science and Technology of China, 
		Hefei 230027, 
		China
}

\begin{document}
\maketitle


\begin{abstract}	
Instability and rupture dynamics of a liquid nano-thread, subjected to external hydrodynamic perturbations, are captured by a stochastic lubrication equation (SLE) incorporating thermal fluctuations via Gaussian white noise. Linear instability analysis of the SLE is conducted to derive the spectra and distribution functions of thermal capillary waves influenced by external perturbations and thermal fluctuations. 
The SLE is also solved numerically using a second-order finite difference method with a correlated noise model.
Both theoretical and numerical solutions, validated through molecular dynamics,
indicate that surface tension forces due to specific external  perturbations overcome the random effects of thermal fluctuations, determining both the thermal capillary waves and the evolution of perturbation growth.
The results also show two distinct regimes: (i) the hydrodynamic regime, where external perturbations dominate, leading to uniform ruptures, and (ii) the thermal-fluctuation regime, where external perturbations are surpassed by thermal fluctuations, resulting in non-uniform ruptures. The transition between these regimes, modelled by a criterion developed from the linear instability theory, exhibits a strong dependence on the amplitudes and wavenumbers of the external perturbations. 
\end{abstract}




\section{Introduction}\label{sec1}

The interfacial dynamics of liquid nano-threads play a crucial role in modern fluid-based techniques, including in-fibre particle production \citep{kaufman2012structured}, fabrication of structures in micro-/nano-electromechanical systems \citep{li2015focused}, and nano-printing \citep{zhang2016micro}. 
Experimentally observing the dynamics at the nanoscale is often challenging, highlighting the significance of modelling and simulation in unraveling the underlying physics.

Classical models describing the macroscale dynamics of liquid threads typically consist of two stages \citep{eggers2008physics}: (i) the linear dynamics of instability and (ii) the nonlinear dynamics leading to rupture.
Theoretical foundations for linear instability were laid by two pioneers: \citet{plateau1873statique} deduced the critical wavelength ($\lambda_\sub{crit}$), below which all interface disturbances decay, and \citet{rayleigh1878instability} identified the fastest-growing mode ($\lambda_\sub{max}$) by applying the normal mode expansion to the axisymmetric Navier-Stokes (NS) equations.
Concerning the nonlinear dynamics, various scaling laws have been developed to describe the final pinch-off stage in three typical scenarios:
the inertial regime \citep{chen1997dynamics, day1998self}, the viscous regime \citep{papageorgiou1995breakup}, and the viscous-inertial regime \citep{eggers1993universal}. Experimental confirmations of these regimes \citep{castrejon2015plethora, lagarde2018oscillating} further indicate that transitions between them are notably intricate.
Building upon these classical models for the Rayleigh-Plateau (RP) instability and rupture, recent studies have employed specific actuations to introduce external perturbations to manipulate the interfacial dynamics of liquid threads/jets at the macroscale, facilitating the generation of uniform droplets \citep{yang2019manipulation, zhao2021uniform,mu2023modulation}.

However, at the nanoscale, classical theories prove inadequate due to the emergence of new physical mechanisms \citep{kavokine2021fluids}. 
One significant factor is thermal fluctuations caused by random molecular motions. Their influence on the interfacial dynamics of liquid threads was first emphasised numerically \citep{koplik1993molecular} and experimentally \citep{shi1994cascade}.
Subsequently, \citet{moseler2000formation} conducted molecular dynamics (MD) simulations of nano-jets to reveal a distinctive ``double-cone'' shape near the rupture point, contrasting the long neck observed at the macroscale.
The discrepancy attributed to thermal fluctuations was elucidated using a stochastic lubrication equation (SLE) derived by applying the slender-body approximation to the governing equations of fluctuating hydrodynamics \citep{landau1987statistical}. 
This approach captures the influence of thermal fluctuations by incorporating stochastic flux terms.
Furthermore, thermal fluctuations have proven to play a vital role in other nanoscale interfacial hydrodynamics, such as fluid mixing \citep{kadau2007importance}, droplet coalescence \citep{perumanath2019droplet}, bounded film flows \citep{zhang2019molecular, zhao2023fluctuation}, and moving contact lines \citep{liu2023thermal}.

For the linear instability of liquid nano-threads, it was initially demonstrated that the classical instability criterion remains applicable at the nanoscale \citep{min2006rayleigh, tiwari2008simulations}. Subsequent research by \citet{gopan2014rayleigh} indicated that thermal fluctuations only affect the dynamics during the last stage of breakup.
However, \citet{mo2016application} contested this, asserting that the growth rates of thermal capillary waves deviate considerably from classical theories.
In a recent study by \citet{zhao2019revisiting}, a framework was developed for modelling the linear instability of interfaces in the presence of thermal fluctuations, named the SLE-RP. At the nanoscale, the SLE-RP shows that the criterion of the classical RP instability can be violated and $\lambda_\sub{max}$ predicted by classical theories is significantly modified (notably becoming time-dependent), recently supported by the numerical simulations for the governing equations of fluctuating hydrodynamics \citep{Barker2023}.

Concerning the nonlinear dynamics, \citet{eggers2002dynamics} first derived a similarity solution from the SLE to describe the nonlinear dynamics of nano-thread rupture. 
This solution successfully replicated the double-cone profile observed by \citet{moseler2000formation} and presented a power law governing the progression of the minimum thread radius to rupture: $h_\sub{min} \sim (t_r-t)^{0.418}$, where $t_r$ represents the rupture time. This power law, distinct from the macroscale counterparts, was subsequently verified experimentally \citep{hennequin2006drop, petit2012break} and numerically \citep{arienti2011many, zhao2020rupture}. However, a recent study by \citet{zhao2020dynamics} challenged this power law, demonstrating its validity only under the condition of ultralow surface tension.

The influence of thermal fluctuations on liquid nano-threads leads to their breakup into droplets with sizes spanning a broad range \citep{gopan2014rayleigh, xue2018effects}, presenting challenges for potential nanoscale applications. Despite extensive investigations into manipulating interfacial dynamics at the macroscale \citep{yang2019manipulation, zhao2021uniform, mu2023modulation}, the exploration of such external perturbations at the nanoscale has been relatively limited.
\citet{fowlkes2012signatures} investigated dewetting of striped liquid films with prescribed perturbations using MD simulations.
The external perturbations with the wavelength $\lambda > \lambda_\sub{crit}$ were found to determine the RP instability of the liquid films and lead to uniform droplets, while perturbations with the wavelength $\lambda < \lambda_\sub{crit}$ proved ineffective in controlling droplet sizes.
But the effects of the thermal fluctuations were not quantitatively analysed in their work.
\cite{shah2019thermal} explored the instability of ultra-thin films driven by both thermal fluctuations and drainage due to the curvature of the initial interface profile (similar to an external perturbation).
The competition between these two effects yields two regimes of the instability: the dimple-dominated regime and the fluctuation-dominated regime.
Notably, only the tangential curvature (for calculating surface tension) was considered in the planar liquid film, while both tangential and circumferential curvatures are crucial in the interfacial dynamics of liquid threads. Surface tension forces due to the latter one serve as the dominant driving force for the instability, distinguishing it significantly from planar liquid films.
Hence, the physics governing the interaction between external perturbations and thermal fluctuations, and their collective impact on the interfacial dynamics of nano-threads, remain uncertain.

In this study, SLE and MD are employed to investigate the effects of external perturbations on the interfacial dynamics of liquid nano-threads. 
The paper is organised as follows. In \S\,\ref{sec2}, we introduce the models of fluctuating hydrodynamics of liquid nano-threads, where the SLE is first presented in \S\,\ref{sec2-1}, followed by analytical solutions of the linearised SLE (\S\,\ref{sec2-2}) and schemes for the numerical solutions of the SLE (\S\,\ref{sec2-3}). 
Subsequently, details of MD simulations are introduced in \S\,\ref{sec3}.
Results in \S\,\ref{sec4} show the influence of the external perturbations and thermal fluctuations on: (i) the thermal capillary wavelengths (\S\,\ref{sec4-1}) and (ii) the evolution of perturbation growth (\S\,\ref{sec4-2}).
Two instability regimes are also defined with boundaries of regime conversions presented in \S\,\ref{sec4-3}.

\section{Fluctuating hydrodynamics modelling \label{sec2}}

In this section, the SLE, as a simple mathematical model, is introduced (\S\,\ref{sec2-1}) to pursue theoretical solutions (\S\,\ref{sec2-2}) and numerical solutions (\S\,\ref{sec2-3}) of the dynamics of liquid nano-threads. 

\subsection{Stochastic lubrication equation \label{sec2-1}}

\begin{figure}
	\centering
	\includegraphics{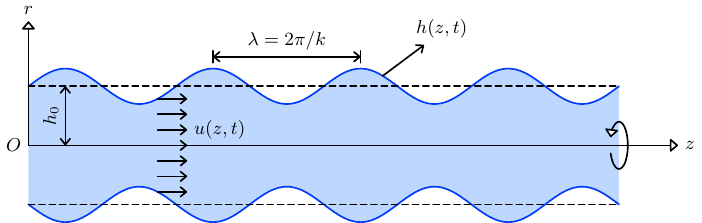}
	
	\caption{
		Schematic of an axisymmetric liquid nano-thread with capillary waves.
	}
	\label{ideal thread}
\end{figure}

We consider an axisymmetric liquid nano-thread in a cylindrical coordinate system, with its axis aligned along the $z$ direction (figure\,\ref{ideal thread}). $h_0$ is the initial radius of the thread. 
The SLE was derived by \citet{moseler2000formation} via applying a lubrication approximation to the axisymmetric fluctuating hydrodynamic equations, allowing the dynamics of the interface to be described by the thread radius $h(z,t)$ and the axial velocity $u(z,t)$.  

To identify the governing dimensionless parameters, we use the following variables as scales of
length, time, velocity and pressure, based on (but not confined to) a balance of inertial and surface-tension forces:
\begin{equation}\label{nondimensionalization}
	h=\frac{h^*}{h_0}, \quad 
	t=\frac{t^*}{\sqrt{\rho h_0^3 \big/ \gamma}}, \quad 
	u=\frac{u^*}{\sqrt{\gamma / \rho h_0}}, \quad 
	p=\frac{p^*}{\gamma / h_0}, \quad 
	N=\frac{N^*}{\sqrt[4]{\gamma \big/ \rho h_0^{5}} },
\end{equation}
where $h^*$, $t^*$, $u^*$ and $p^*$ respectively indicate dimensional thread radius, time, velocity and pressure. The variables without asterisks represent corresponding dimensionless ones (note that  the dimensional material parameters are not given asterisks).
$\rho$ is the density and $\gamma$ the surface tension.
To model thermal fluctuations, a stochastic term $N(z,t)$ is introduced, standing for a Gaussian white noise that obeys the fluctuation-dissipation theorem.
Its mean and autocovariance are respectively calculated as
\begin{subequations}\label{gwn}
	\begin{empheq}[left={\empheqlbrace}]{alignat=1}
		\left\langle N(z,t) \right\rangle &= 0, \\
		\left\langle N(z,t)N(\acute{z}, \acute{t}) \right\rangle &= \delta(z-\acute{z})\delta(t-\acute{t}),
	\end{empheq}
\end{subequations}
where $\delta$ represents a unit impulse function and $	\left\langle \cdot \right\rangle $ denotes ensemble averages.
$\acute{t}$ and $\acute{z}$ could be infinitesimally close to the original ones in time or space.
The dimensionless SLE can be written as
\begin{subequations}\label{dimensionless SLE}
	\begin{empheq}[left={\empheqlbrace}]{alignat=1}
		\frac{\partial u}{\partial t}&=-u u^{\prime}-p^{\prime}+3 \ohn \frac{\left(h^{2} u^{\prime}\right)^{\prime}}{h^{2}}+\sqrt{\frac{6\ohn}{\pi}}\thn \frac{(h N)^{\prime}}{h^{2}}, \label{dimensionless SLE 1} \\
		\frac{\partial h}{\partial t}&=-u h^{\prime}-\frac{u^{\prime} h}{2}, \label{dimensionless SLE 2}
	\end{empheq}
\end{subequations}
where the superscript primes denote partial derivatives with respect to $z$.
In equation\,\eqref{dimensionless SLE}, one dimensionless quantity is the Ohnesorge number, which relates the viscous forces to inertial and surface-tension forces, i.e. $\ohn = \mu \big/ \sqrt{\rho \gamma h_0}$. Here, $\mu$ is the dynamic viscosity.
Another dimensionless quantity is thermal-fluctuation number $\thn = l_\sub{T} / h_0$, representing the relative intensity of interfacial fluctuations.
Here, $l_\sub{T}= \sqrt{\kb T / \gamma}$ is the characteristic thermal-fluctuation length, with $\kb$ being the Boltzmann constant and $T$ the temperature.
When $\thn=0$, the SLE reduces to the deterministic lubrication equation (LE) proposed by \citet{eggers1994drop}.
Additionally, the full Laplace pressure in \eqref{dimensionless SLE} is determined from the principal curvatures
\begin{equation}\label{laplace pressure}
	p=\frac{1}{h \sqrt{1+\left(h^{\prime}\right)^{2}}}-\frac{h^{\prime \prime}}{\left[1+\left(h^{\prime}\right)^{2}\right]^{\frac{3}{2}}}.
\end{equation}

\subsection{Linear instability analysis \label{sec2-2}}

For the linear instability, we take $h(z,t) = 1 + \tilde{h}(z,t)$ and assume that $\tilde{h}(z,t) \ll 1$. Substituting this into equations \eqref{dimensionless SLE} and ignoring higher order terms give the linearised SLE
\begin{equation}\label{LSLE}
	\frac{\partial^{2} \tilde{h}}{\partial t^{2}}-3\ohn\frac{\partial\tilde{h}^{\prime\prime}}{\partial t}+\frac{1}{2}\left(\tilde{h}^{\prime\prime}+\tilde{h}^{\prime\prime\prime\prime}\right) = -\sqrt{\frac{3\ohn}{2\pi}}\thn N^{\prime\prime}.
\end{equation}
A Fourier transform within $[0, L]$ is then applied for equation\,\eqref{LSLE} to give a second-order ordinary differential equation
\begin{equation}\label{transformed LSLE}
	\frac{\d^2 H}{\d t^2}+3\ohn k^2\frac{\d H}{\d t}+\frac{k^4-k^2}{2} H=\sqrt{\frac{3\ohn}{2\pi}}\thn\, k^2 \xi,
\end{equation}
where
\begin{equation} 
	H(k,t) = \int_{0}^{L}\tilde{h}(z,t) \e^{-\i kz} \dd z, \quad \xi(k,t) = \int_{0}^{L}N(z,t) \e^{-\i kz} \dd z. 
\end{equation}
Here, $k$ is the wavenumber. 
The transformed variable $H$ represents the spectrum of the thermal capillary waves of the interfaces.  Its final solution is expressed as follows (see Appendix\,\ref{appc} for derivation)
\begin{equation}\label{Hrms}
	\left|H\right|_\sub{rms} = \sqrt{\left\langle \left|\HL\right|^2 \right\rangle + \left\langle \left|\HF\right|^2 \right\rangle},
\end{equation}
where
\begin{subequations}\label{HLE and Hf}
	\begin{empheq}[left={\empheqlbrace}]{alignat=2}
		&\left\langle \left|\HL\right|^2 \right\rangle &&= \left|H_\sub{i}\right|^2\e^{-at} \left[ \cosh\left( \frac{bt}{2} \right) + \frac{a}{b}\sinh\left( \frac{bt}{2} \right) \right]^2, \label{HLE}\\
		&\left\langle \left|\HF\right|^2 \right\rangle &&= \frac{3L\ohn}{\pi}\thn^2 k^4\frac{a^2-b^2-a^2 \cosh (b t)-a b \sinh (b t)+b^2 \e^{a t}}{a b^2\left(a^2-b^2\right) \e^{a t}}. \label{Hf}
	\end{empheq}
\end{subequations}
Here, the subscript ``$\sub{rms}$'' represents root mean square, $a=3\ohn k^2$ and $b=\sqrt{\left(9\ohn^2 -2\right)k^4+2k^2}$.
The solution is linearly decomposed into two terms: the hydrodynamic component $\HL$ and the thermal-fluctuation component $\HF$.
$\HL$ is the solution to the homogeneous form of equation\,\eqref{transformed LSLE}, representing the classical (deterministic) RP instability;
$\HF$ arises from solving the full form of equation\,\eqref{transformed LSLE} with zero initial disturbances, representing the fluctuation-drive instability.
In equation\,\eqref{HLE}, $H_\sub{i}$ is the initial spectrum, determined by the capillary waves at $t=0$.
With initial perturbation waves, $\tilde{h}(z, 0)=A_0\sin \left(k_0 z\right)$, we have
\begin{equation}\label{Hi}
	\left| H_\sub{i} \right| = \left|\int_{0}^{L}A_0\sin\left(k_0z\right)\e^{-\i kz}\dd z\right| = 2A_0k_0\left| \frac{\sin\left(kL/2\right)}{k^2-k_0^2} \right|.
\end{equation}

\subsection{Numerical scheme \label{sec2-3}}

In this work, we solve the SLE numerically with periodic boundary conditions using the MacCormack method \citep{maccormack2003effect}, a simple second-order explicit finite difference scheme in both time and space.
At each time level, the solution is represented by two arrays: $\{h_i \}^M_{i=1}$ and $\{u_i\}^M_{i=1}$, where $M$ denotes the number of mesh points. The time-derivative terms are approximated as $\left(h_i^{t+1}-h_i^t\right) \big/ \rmDelta t$ and $\left(u_i^{t+1}-u_i^t\right) \big/ \rmDelta t$.
The numerical method follows a two-step process, beginning with a predictor step

\begin{equation}
	\begin{pmatrix}
		\bar{u}^{t+1}_i \\[4pt]
		\bar{h}^{t+1}_i 
	\end{pmatrix} 
	=
	\begin{pmatrix}
		u^t_i \\[4pt]
		h^t_i 
	\end{pmatrix}
	+ \bm{D}\left(u^t_i, h^t_i\right) \rmDelta t,
\end{equation}
and a corrector step
\begin{equation}
	\begin{pmatrix} 
		u^{t+1}_i \\[4pt]
		h^{t+1}_i    
	\end{pmatrix} 
	=
	\begin{pmatrix} 
		u^t_i \\[4pt]
		h^t_i    
	\end{pmatrix} 
	+ \frac{\rmDelta t}{2} \left[ \bm{D}\left(u^t_i, h^t_i\right)+\bar{\bm{D}}\left( \bar{u}^{t+1}_i, \bar{h}^{t+1}_i \right) \right],
\end{equation}
where $\bar{u}_i^{i+1}$ and $\bar{h}_i^{t+1}$ denote the ``provisional" values at time level $t+1$, and $\bm{D}$ encompasses all the partial spatial derivative terms on the right-hand side (expressions for $\bm{D}$ are listed in Appendix\,\ref{appa}).

For the stochastic term $N(z,t)$, the autocovariance of uncorrelated fluctuations can be numerically approximated by a 2D rectangular (boxcar) function, non-zero over a time step ($\rmDelta t$) and grid spacing ($\rmDelta z$), expressed as $N(z,t) \approx N_i^t  = \mathcal{N} \big/ \sqrt{\rmDelta t \rmDelta z}$.
Here, $\mathcal{N}$ signifies computer-generated random numbers, following a normal distribution with zero mean and unit variance.
However, this model has been shown prone to numerical instability for the SLE \citep{zhao2020dynamics}, problems that are exacerbated as $\rmDelta z$ and $\rmDelta t$ become smaller and the amplitude of noise becomes larger.

To establish a robust numerical scheme, we adopt the methodology introduced by \citet{zhao2022fluctuation}, combining a spatially and temporally correlated noise model. This integration enforces correlations in the noise beneath the spatial correlation length $l_\sub{c}$ and the temporal correlation length $t_\sub{c}$. The uncorrelated behavior is then approximated by taking the limit of these lengths approaching zero, ensuring their numerical resolution remains accurate throughout the limiting process.
The stochastic term $N(z,t)$ is expanded using separation of variables in the $Q$-Wiener process $W(z,t)$ \citep{grun2006thin,diez2016metallic}, as follows
\begin{equation}
	N(z,t) = \frac{\partial W(z,t)}{\partial t} 
	= \sum^{\infty}_{q = -\infty} \chi_q\, \dot{c}_q(t)\, g_q(z) \,.
\end{equation}
Here, $\chi_q$ represents the eigenvalues of the correlation function $F_\sub{c}$,
\begin{equation}
	\chi_q = \int_{0}^{L} F_\sub{c}(z) \e^{-2\pi\i q z/L} \dd z,
\end{equation}
where $q$ represents an integer sequence.
The expressions for $F_\sub{c}$, $g_q$ and other details regarding this model can be found in Appendix\,\ref{appb}.
The coefficient $\dot{c}_q(t)$, representing a temporally correlated noise process, is modelled using a straightforward linear interpolation between uncorrelated random noise at the endpoints of the temporal correlation interval, as proposed by \citet{zhao2020dynamics}. Therefore, the final discretised expression of the noise term becomes
\begin{equation}
	N^t_i= \frac{1}{\sqrt{t_\sub{c}}}\sum^{\frac{M+1}{2}}_{q=-\frac{M+1}{2}} \chi_q\, \mathcal{N}_q\, g_q(z) \,.
\end{equation}

In this work, we set the dimensionless grid size $\rmDelta z= 0.03$ and time step $\rmDelta t = 5 \times 10^{-6}$ (the dimensional parameters corresponding to MD simulations are $\rmDelta z^* = 0.11 \, \mathrm{nm}$ and $\rmDelta t^* = 0.13 \, \mathrm{fs}$ respectively).
The correlation lengths $l_\sub{c} = 5 \rmDelta z=0.15$ and $t_\sub{c} = 10 \rmDelta t = 5 \times 10^{-5}$. 	
A discussion of the influence of $l_\mathrm{c}$ and $t_\mathrm{c}$ is presented in Appendix\,\ref{app_tc_lc}.
The boundary conditions are set as periodic.

\section{Molecular dynamics \label{sec3}}
\begin{figure}
	\centering
	\includegraphics{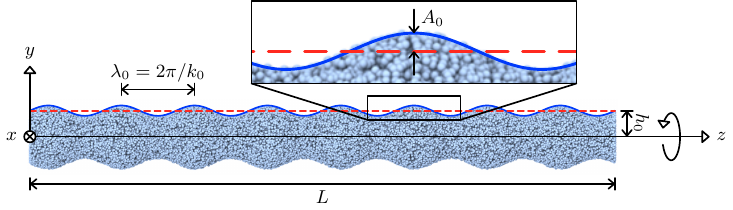}
	
	\caption{
		Initial configuration for MD simulations: a liquid nano-thread with an external perturbation.
	}
	\label{MD thread}
\end{figure}

The MD simulations of this work are performed using the open source package LAMMPS \citep{thompson2022lammps}.
The simulation box ($40\,\mathrm{nm} \times 40\,\mathrm{nm} \times 432 \,\mathrm{nm}$ in the $x$, $y$ and $z$ directions, respectively) has periodic boundary conditions imposed in all directions.
A nano-thread of water, initially with a radius of $h_0=3.6\,\mathrm{nm}$, is positioned at the center of the simulation domain.
The thread length is equal to the length of the simulation box in the $z$ direction, i.e the dimensionless thread length $L=120$.
External perturbations are introduced through a sinusoidal function, $h(z,t) = h_0+A_0\sin\left( k_0 z \right)$, where $A_0$ denotes the initial amplitude and $k_0$ represents the (angular) wavenumber (figure\,\ref{MD thread}). 
Initial configurations with various external perturbations are generated from equilibrium simulations of a liquid bulk in the canonical (NVT) ensemble, employing the Nos\'e-Hoover thermostat at $T = 300 \, \mathrm{K}$. 
The interactions between water molecules are described by a coarse-grained force field, the mW potential \citep{molinero2009water}. 
The final density of the bulk at equilibrium is $997\,\mathrm{kg}\big/\mathrm{m}^{3}$.
Considering  the ultralow density of vapours predicted by the mW model, the nano-threads are simulated in a vacuum environment with a timestep of $2.5 \, \mathrm{fs}$. The canonical ensemble with the Nos\'e-Hoover thermostat is employed again to keep $T = 300 \, \mathrm{K}$.

To determine $\ohn$ and $\thn$, viscosity and surface tension of the liquid are required to be extracted from MD. Here, we employ the Green-Kubo relation \citep{green1954markoff, kubo1957statistical} to calculate the dynamic viscosity
\begin{equation}
	\mu=\frac{V}{\kb T} \int_0^{\infty} \left\langle p_{ij}(t) p_{ij}(0) \right\rangle \dd t,\quad i\neq j,
\end{equation}
where $V$ is the volume of a liquid bulk, $p_{ij}$ the off-diagonal elements of the pressure tensor and $\left\langle p_{ij}(t) p_{ij}(0) \right\rangle$ the autocorrelation function of $p_{ij}$. In addition, a liquid layer lying in the $x$-$y$ plane is used to estimate the surface tension. Resorting to pressures on the two free surfaces, we have this expression \citep{kirkwood1949statistical}
\begin{equation}
	\gamma=\int_{-\infty}^{\infty}\left[p_\sub{n}(z)-p_\sub{t}(z)\right]\dd z,
\end{equation}
where the normal and tangential pressure components are defined as $p_\sub{n} = p_{zz}$ and $p_\sub{t} = \left( p_{xx}+p_{yy} \right) \big/ 2$, respectively. Finally, we have $\mu = 3.14\times 10^{-4} \, \mathrm{Pa\cdot s}$ and $\gamma = 6.5\times 10^{-2} \, \mathrm{N/m}$ at $T = 300 \, \mathrm{K}$, leading to $\ohn = 0.65$ and $\thn = 0.07$.

\section{Results and discussions}\label{sec4}
In this section, the analytical and numerical solutions of the SLE are validated by MD results, showing the influence of the thermal fluctuations and external perturbations on the thermal capillary waves  (\S\,\ref{sec4-1}) and evolution of perturbation growth (\S\,\ref{sec4-2}). 
Two instability regimes are also defined with boundaries of regime conversions presented in \S\,\ref{sec4-3}.

\subsection{Thermal capillary waves}\label{sec4-1}

To examine the theoretical solution in \S\,\ref{sec2-2}, we conduct MD simulations on long threads with  various wavenumbers $k_0$ and amplitudes $A_0$: case\,1 ($A_0=0$), case\,2 ($A_0=0.1$, $k_0=\pi/6$) and case\,3 ($A_0=0.2$, $k_0=2\,\pi/5$).
For each case, 30 independent MD simulations (realisations) are performed to gather statistics. 

\begin{figure}
	\centering
	\includegraphics{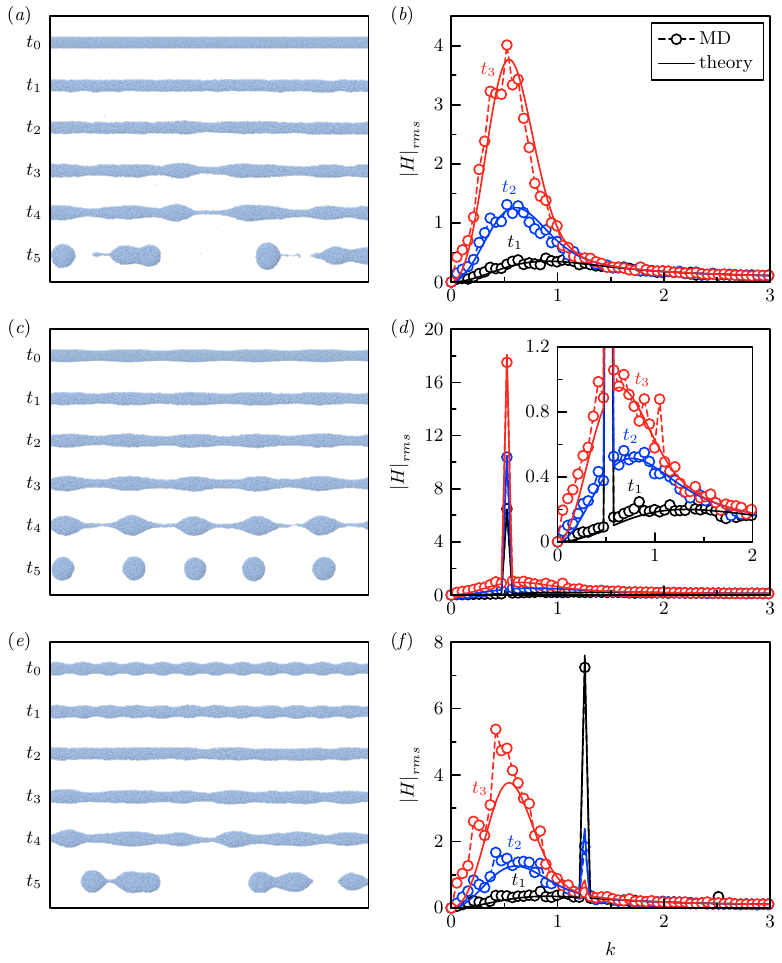}
	
	\caption{
		The left panel shows the MD snapshots; the right panel shows the comparisons between theoretical (solid lines) and numerical (dashed lines with circles) spectra in cases\,1--3. 
		(\textit{a},\textit{b}) Results of case\,1 at six instants: 0.00, 3.27, 10.76, 18.24, 21.98 and 29.93. 
		(\textit{c},\textit{d}) Results of case\,2 at six instants: 0.00, 1.40, 5.14, 8.88, 15.90 and 22.45. 
		(\textit{e},\textit{f}) Results of case\,3 at six instants: 0.00, 3.27, 10.76, 18.24, 23.85 and 32.73. 
		The liquid threads illustrated here are all truncated at $z = L/2$.
	}
	\label{MD and spectra}
\end{figure}

The left panel of figure\,\ref{MD and spectra} illustrates the MD snapshots of cases\,1--3.
Specifically, case\,1 represents a situation with no external perturbations, while cases\,2 and 3 involve different external perturbations.
In case\,1 (figure\,\ref{MD and spectra}\,\textit{a}), perturbations arising from thermal fluctuations grow over time, generating significant capillary waves and eventually lead to the final rupture at $t_4$. 
Since this fluctuation-driven instability is naturally stochastic, the liquid thread break up into non-uniform droplets.
In contrast, the external perturbation in case\,2 grows, despite being disturbed by the thermal fluctuations, and ultimately leads to a uniform rupture similar to the macroscale cases actively controlled (figure\,\ref{MD and spectra}\,\textit{c}).
The external perturbation with a larger wavenumber (compared to $k_0$ in case\,2) decays rapidly and is then overwhelmed by fluctuation-drive perturbations (figure\,\ref{MD and spectra}\,\textit{e}), leading to an irregular breakup pattern similar to that in case\,1. 

The phenomena presented above can be further explained quantitatively by the spectra  in the right panel of figure\,\ref{MD and spectra}. 
Following the approach used by \citet{zhao2021influence}, the profile $h(z,t)$ in each MD realisation is extracted from axially distributed annular bins based on a threshold value of particle number density.
A discrete Fourier transform is applied to $h(z,t)$ to get the spectra.
We then ensemble average the spectra at each instant over the realisations and take the square root to produce the numerical spectra of cases\,1--3 in figure\,\ref{MD and spectra} (dashed lines with circles). 
Good agreement with the theoretical model for the spectra can be found for all the cases, confirming the validity of equation\,\eqref{Hrms}.

The spectra of case\,1 (figure\,\ref{MD and spectra}\,\textit{b}) display a modal distribution with a certain bandwidth at each time instant, explaining why the liquid thread exhibits a non-uniform breakup.
In cases\,2 and 3, the external perturbations lead to initial spikes in the spectra, representing the initial conditions of the hydrodynamic component $\HL$, modelled by $H_\sub{i}$ of equation\,\eqref{Hi}.
Note that equation\,\eqref{Hi} can cause ``spectral leakage'' \citep{proakis1996digital}, which leads to noise in the initial spectra. 
To avoid this problem and compare with the results from the discrete Fourier transform, further processing on equation\,\eqref{Hi} is required (see Appendix\,\ref{appd} for details).
The spike in case\,2 increases rapidly and indicates that the hydrodynamic component predominates the entire dynamics, resulting in the formation of uniform droplets after the rupture. However, the spike in case\,3 decays drastically and is overwhelmed by the fluctuation modes, denoting that thermal fluctuations re-dominate the instability.
The difference between cases\,2 and 3 can be explained by the classical RP theory \citep{plateau1873statique,rayleigh1878instability}, where perturbations of short wavelength $\lambda < \lambda_\sub{crit}$ would dissipate.

\begin{figure}
	\centering
	\includegraphics{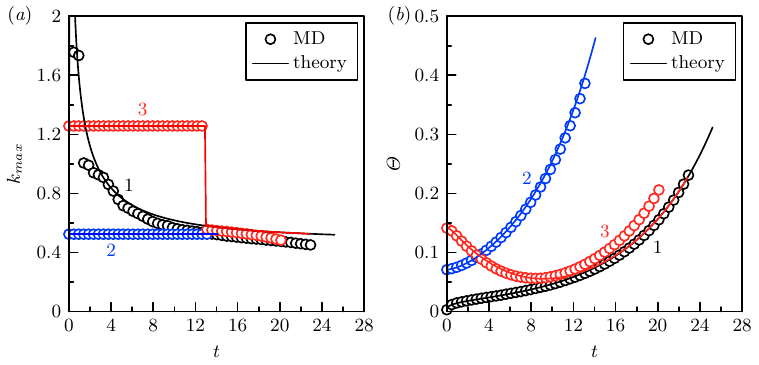}

	\caption{
		MD (circles) and theoretical (solid lines) results of temporal evolutions of (\textit{a}) dominant wavenumber $k_\sub{max}$ and (\textit{b}) surface roughness $\varTheta$ in cases\,1 (black), 2 (blue) and 3 (red).
	}
	\label{kmax and roughness}
\end{figure}

The dominant wavenumbers $k_\sub{max}$ of the instability are also extracted from the peaks of spectra, illustrated in figure\,\ref{kmax and roughness}(\textit{a}).
For case\,1, $k_\sub{max}$ decreases monotonically to a constant, which has been pointed out by \citet{zhao2019revisiting}. 
Since the external perturbation dominates the instability of case\,2, its $k_\sub{max}$ maintains an invariant value, i.e. $k_\sub{max} = k_0=\pi/6$.
In case\,3, $k_\sub{max}$ remains equal to $k_0=2\pi/5$ at the early stage.
When the peak of $k_0$ is surpassed by the instability modes due to the thermal fluctuations,  $k_\sub{max}$ return to the trajectory observed in case\,1.

Moreover, we define the evolution of surface roughness $\varTheta(t)$ via integrating the square of $\tilde{h}(z,t)$ over the entire spatial domain to measure the development of the thermal capillary waves.
According to the Parseval's theory, $\varTheta(t)$ can be expressed as
\begin{align}\label{roughness}
	\Theta^2 &= \frac{1}{L} \left\langle \int_{0}^{L}\tilde{h}^2 \dd z \right\rangle  = \frac{1}{\pi L}\int_0^{\infty}\left|H\right|_\sub{rms}^2 \dd k .
\end{align}
Since $\left|H\right|_\sub{rms}$ consists two components, the surface roughness is also divided into
\begin{equation}
	\Theta^2 = \Theta_\sub{LE}^2 + \Theta_\sub{fluc}^2 = 
	\frac{1}{\pi L}\left( \int_0^{\infty}\left\langle \left|\HL\right|^2 \right\rangle\dd k 
	+\int_0^{\infty}\left\langle \left|\HF\right|^2 \right\rangle\dd k
	\right).
\end{equation} 
Figure\,\ref{kmax and roughness}(\textit{b}) illustrates the evolution of the roughness versus time in cases\,1--3. 
The roughness $\varTheta$ increases constantly with time in both cases\,1 and 2. 
Case\,2 exhibits a higher initial growth rate due to external perturbations. In case\,3, the surface roughness initially decreases due to the dissipation of initial hydrodynamic perturbation, and then increases driven by thermal fluctuations.

The spectra above only present the interfacial dynamics in the frequency domain, i.e. $\left|H\right|_\sub{rms}(k,t)$. 
To gain a better understanding of dynamics in the spatial domain, we propose a distribution function of the perturbation amplitudes, $P(\hat{h})$, where $\hat{h}$ is introduced to represent a possible value of the random perturbation amplitudes $\tilde{h}$.
The perturbations at the linear stage can be divided into two independent components: $\tilde{h} = \tilde{h}_\sub{LE} + \tilde{h}_\sub{fluc}$, where $\tilde{h}_\sub{LE}$ represents waves generated by the classical RP instability and $\tilde{h}_\sub{fluc}$ accounts for waves from thermal fluctuations.
So $P(\hat{h})$ can be modelled by the convolution of the probability distributions of each components \citep{rice2007mathematical}, expressed as
\begin{equation}\label{convolution}
	P= P_\sub{LE} \otimes 
	P_\sub{fluc} \,.
\end{equation}
Here ``$\otimes$'' denotes convolution.

To get the expression of $P_{LE}$, we introduce the cumulative distribution function $F_{\tilde{h}}$ 
\begin{equation}
	F_{\tilde{h}} ( \hat{h} ) = \int_{-\infty}^{\hat{h}} P_\sub{LE} \dd \tilde{h}\,.
\end{equation}
Based on the classical RP instability, $\hat{h} =  A(t)\sin\left( k_0 \hat{z} \right)$, where $A$ grows or decays exponentially from the initial value $A_0$.
So $\hat{h}$ and $\hat{z}$ have a one-to-one functional relationship and are piecewise monotonic.
It is easy to get an inverse function, $\hat{z} = \mathrm{arcsin}(\hat{h}/A)/k_0$.
According to the approach of the distribution function transformation  \citep{edition2002probability}, $F_\sub{\tilde{h}}(\hat{h})$ is equal to the cumulative distribution function of $\hat{z}$, $F_z(\hat{z})$. When $z$ follows a uniform distribution, we have
\begin{equation}\label{LE distribution}
	P_\sub{LE}( \hat{h} ) 
	= \frac{\d F_\sub{z}\left( \hat{z} \right)}{\d \hat{h}} 
	= \frac{\d F_z}{\d \hat{z}} \left| \frac{\d \hat{z}}{\d \hat{h}} \right| 
	= \frac{1}{\pi\sqrt{A^2 - \hat{h}^2}}\,,
\end{equation}
where $\hat{h} \in (-A, A) $. Note that the final expression of $P_\sub{LE}$ is independent of the wavenumbers ($k_0$) of the initial perturbations.

\begin{figure}
	\centering
	\includegraphics{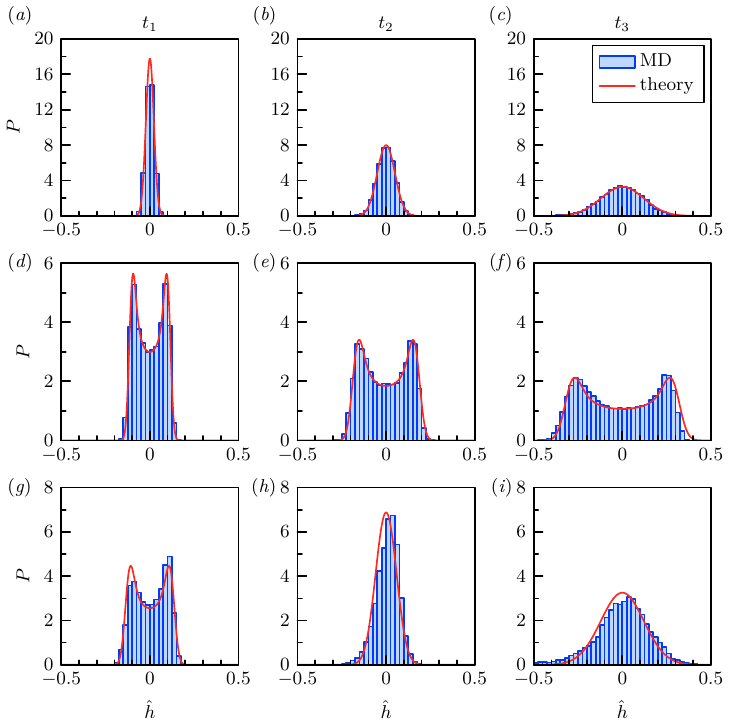}
	
	\caption{
		Temporal evolutions of $P$ from MD simulations (histograms) and the theory (red solid lines) in cases\,1--3.
		(\textit{a},\textit{b},\textit{c}) Results of case\,1 at three instants: 3.27, 10.76 and 18.24.
		(\textit{d},\textit{e},\textit{f}) Results of case\,2 at three instants: 1.40, 5.14 and 8.88.
		(\textit{g},\textit{h},\textit{i}) Results of case\,3 at three instants: 3.27, 10.76 and 18.24.
	}
	\label{distribution}
\end{figure}

Additionally, it is challenging to pursue a theoretical model of $P_\sub{fluc}$ mainly due to the complexity of equation\,\eqref{LSLE}.
So we extract numerically-predicted $P_\sub{fluc}$ from the MD simulations of case\,1, where $P_{LE}$ can be neglected.
We collect all the values of $\tilde{h}(z)$ from 30 realisations and then plot the numerical distributions of $\tilde{h}$ by the histograms in figures\,\ref{distribution}(\textit{a}--\textit{c}).
Promisingly, $\tilde{h}_\sub{fluc}$ is observed to follow a Gaussian distribution with a mean of zero. 
Moreover, the standard deviation of $\tilde{h}_\sub{fluc}$ is equal to $\varTheta_\sub{fluc}$ in equation\,\eqref{roughness}. Therefore, we have a ``semi-theoretical" model
\begin{equation}\label{fluctuation distribution}
	P_\sub{fluc}(\hat{h}) = \frac{1}{\sqrt{2\pi}\Theta_\sub{fluc}}\exp\left( -\frac{\hat{h}^2}{2\Theta_\sub{fluc}^2} \right).
\end{equation}
From a theoretical perspective, $\hat{h} \in (-\infty, \infty)$ in \eqref{fluctuation distribution}. However, $\hat{h}$ typically falls within $[-3 \Theta_\sub{fluc}, 3 \Theta_\sub{fluc}]$, accounting for a $99.7\%$ confidence interval. Here, $\left| 3 \Theta_\sub{fluc} \right| < 1$, ensuring that the perturbation amplitude is always smaller than the thread radius.
The model is validated by the good agreement between the its predictions and MD results in figure\,\ref{distribution}(\textit{a}--\textit{c}).
A more rigorous derivation of $P_\sub{fluc}$ involves pursuing the Fokker-Planck equation of \eqref{LSLE}, which would be a subject of our future research.
Combining equations\,\eqref{convolution}, \eqref{LE distribution} and \eqref{fluctuation distribution} gives us the final theoretical expression of $P(\hat{h})$.

Figures\,\ref{distribution}(\textit{d}--\textit{i}) compare the numerical results from MD simulations with the theoretical distributions predicted by \eqref{convolution} for cases\,2 and 3.
 In case\,2, the distribution function maintains a bimodal curve, signifying that the interface can largely preserve the sinusoidal feature. Similar to the trend in case\,1, the two spikes also propagate outward as the thermal capillary waves develop. In case\,3, the initial spikes dissipate and ultimately merge into a Gaussian curve, aligning with the observations in figure\,\ref{kmax and roughness}.

	\subsection{Evolution of perturbation growth}\label{sec4-2}
	
	Besides the distribution of wavelengths (wavenumbers) investigated in \S\,\ref{sec4-1}, the growth of the perturbations, particularly in the cases with uniform rupture, is explored in this section. 
	MD simulations are performed on long threads with  an initial amplitude $A_0=0.2$ and various wavenumbers: $k_0=\pi/15$ for case\,4, $k_0=\pi/12$ for case\,5 and $k_0=2\pi/15$ for case\,6.
	Additionally, numerical simulations for the SLE are also conducted to compare with the MD results and provide deeper insights into the evolution of perturbation growth.
	
		\begin{figure}
		\centering
		\includegraphics[width=0.9\textwidth]{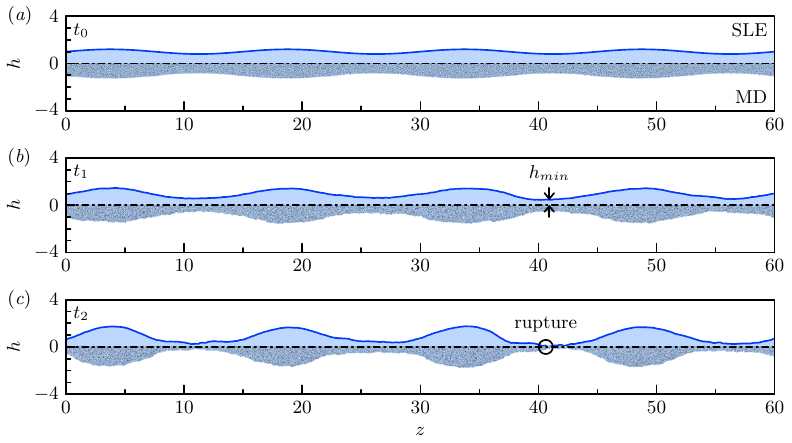}
		
		\caption{
				Evolution of interface profiles extracted from the selected SLE (top) and MD (bottom) simulations. The minimum thread radius $h_\sub{min}$ and the rupture point are marked in (\textit{b},\textit{c}), respectvely.
		}
		\label{com_SLE_MD}
	\end{figure}
	
	Figure\,\ref{com_SLE_MD} displays two selected realisations at three instants from both the MD and SLE simulations of case\,6.
	Though the SLE predictions deviate slightly from the ``double-cone" profile documented by \citet{moseler2000formation}, they agree with the MD results well qualitatively.
	To study the evolution of the perturbations, we focus on the temporal evolution of the minimum (over $z$) thread radius, i.e. $h_{min}(t)$.
	To get statistics, we conduct multiple independent realisations: 30 for MD and 100 for the SLE.

\begin{figure}
		\centering
		\includegraphics{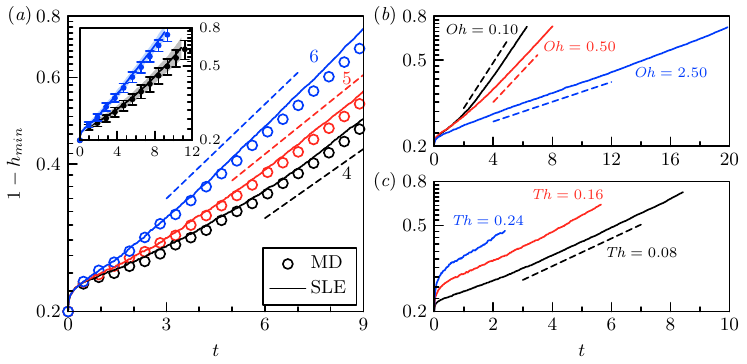}
		
		\caption{ 
				Evolution of the perturbation growth at the linear stage:
				(\textit{a}) the ensemble-averaged SLE predictions (solid lines) and MD results (circles) for cases\,4 (black), 5 (red) and 6 (blue). Here $\ohn = 0.65$ and $\thn = 0.07$. The dashed lines represent the growth rates predicted by the dispersion relation \eqref{LE_dispersion} for the three specific wavenumbers in cases\,4--6: $\omega = 0.108$ for $k_0 = \pi/15$ (black), $\omega = 0.124 $ for $k_0 = \pi/12$ (red) and $\omega = 0.148 $ for $k_0 = 2\pi/15$ (blue). The error bars and shadows in the inset represent one standard
				deviation (either side of the mean) for MD and the SLE, respectively.
				(\textit{b}) The SLE predictions for three values of $\ohn$:  $\omega = 0.244$ for $\ohn=0.1$ (black), $\omega = 0.168 $ for $\ohn=0.5$ (red) and $\omega = 0.053 $ for $\ohn=2.5$ (blue). Here $\thn=0.07$, $k_0 = 2\pi/15$ and $A_0 = 0.2$.
				(\textit{c}) The SLE predictions for three values of $\thn$: 0.08 (black), 0.16 (red) and 0.24 (blue). $\omega = 0.148$ for $\ohn=0.165$ (black). Here $k_0 = 2\pi/15$ and $A_0 = 0.2$.
		}
		\label{linear_SLE_MD}
	\end{figure}

	According to the classical theory \citep{eggers2008physics}, the growth of the perturbations can be divided into the linear and nonlinear stages.
	Figure\,\ref{linear_SLE_MD}(\textit{a}) illustrates the ensemble-averaged perturbation growth ($1-h_{min}$) at the linear stage, extracted from both the numerical solutions of the SLE and the MD results.
	For cases\,4--6, good agreement is observed at all instants for the both mean values and standard deviations (from the thermal fluctuations), further validating the numerical solutions of the SLE.  
	Interestingly, the perturbation is found to grow exponentially, approximately following the relation $1-h_{min} \sim \e^{\omega t}$.
	Despite the presence of the thermal fluctuations, the growth rate $\omega$ is close to the analytical results (dashed lines in figure\,\ref{linear_SLE_MD}\textit{a}), predicted by the dispersion relation of the LE \citep{eggers1994drop}
	\begin{equation}\label{LE_dispersion}
		\omega = \frac{k}{2}\sqrt{9\ohn^2k^2 + 2(1-k^2)} - \frac{3}{2}\ohn k^2.
	\end{equation}
	These observations further explain the occurrence of uniform breakup.
	The surface tension forces, induced by the external perturbations with specific wavenumbers, overcome the random effects due to thermal fluctuations, determining  the final form of the thermal capillary waves.
	Moreover, the influence of $\ohn$ and $\thn$ is investigated using the SLE solver with $k_0$ and $A_0$ from case\,6. We set $\thn=0.07$ in figure\,\ref{linear_SLE_MD}(\textit{b}) and $\ohn=0.65$ in figure\,\ref{linear_SLE_MD}(\textit{c}). Figure\,\ref{linear_SLE_MD}(\textit{b}) shows that the growth rates of the perturbations, which decline with increasing $\ohn$, agree well with the predictions of equation\,\eqref{LE_dispersion}, further confirming the dominant roles of the surface tension forces induced by the external perturbations.
	However, when $\thn$ increases, the growth rate deviates from the predictions of equation\,\eqref{LE_dispersion}, indicating that thermal fluctuations regain a significant role. Notably, each realisation evolves over different time periods, so the ensemble average can only account for the shortest time across all realisations. When thermal fluctuations become crucial, the variance in evolution time is larger (i.e. the minimum of rupture time is smaller), hence the trajectory for the case with $\thn=0.24$ is quite short.
	
	\begin{figure}
		\centering
		\includegraphics{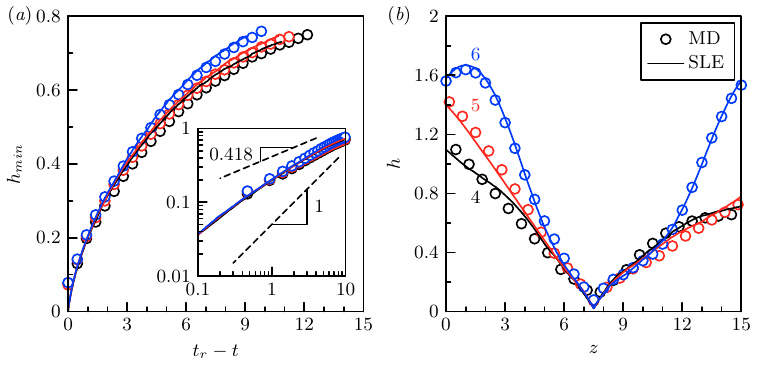}
		
		\caption{
				(\textit{a}) Minimum thread radius ($h_{min}$) against time to rupture $(t_r - t)$ for cases\,4 (black), 5 (red) and 6 (blue): comparison between ensemble-averaged MD results (circles) and SLE calculations (solid lines). The inset illustrates $h_{min}(t_r-t)$ on a logarithmic scale. 
				The two dashed lines represent two power laws of similarity solutions for the rupture dynamics \citep{eggers1993universal,eggers2002dynamics}.
				(\textit{b}) Ensemble-averaged rupture profiles of cases\,4 (black), 5 (red) and 6 (blue):  comparison between ensemble-averaged MD results (circles) and SLE calculations (solid lines).
		}
		\label{nonlinear_SLE_MD}
	\end{figure}
	
	To investigate the nonlinear evolution near rupture, $h_{min}$ extracted from simulations is plotted against time to rupture, $t_r-t$, shown in figure\,\ref{nonlinear_SLE_MD}(\textit{a}).
	The nonlinear dynamics of cases\,4--6 is found to be nearly identical as approaching the rupture point, indicating that external perturbations do not affect the rupture dynamics despite their significant impacts on the evolution at the linear stage. 
	Additionally, the inset of figure\,\ref{nonlinear_SLE_MD}(\textit{a}) suggests that a power law might govern the progression of the minimum thread radius to rupture: $h_{min} \sim (t_r-t)^\alpha$. 
	However, the power law does not satisfy either the thermal-fluctuation-dominated power law, $\alpha=0.418$ \citep{eggers2002dynamics}, or the surface-tension-dominated one, $\alpha=1$ \citep{eggers1993universal}. Instead, it lies between the two, indicating that both fluctuations and surface tension forces contribute to the dynamics during the rupture stage.
	Additionally, figure\,\ref{nonlinear_SLE_MD}(\textit{b}) shows the ensemble-averaged rupture profiles of cases\,4--6.
	The overall interface shapes varies due to  the influence of external perturbations with different wavelengths, whereas profiles near the rupture overlap, further supporting the conclusion in figure\,\ref{nonlinear_SLE_MD}(\textit{a}) that external perturbations do not impact the rupture dynamics.

\subsection{Regime transition}\label{sec4-3}

\begin{figure}
	\centering
	\includegraphics{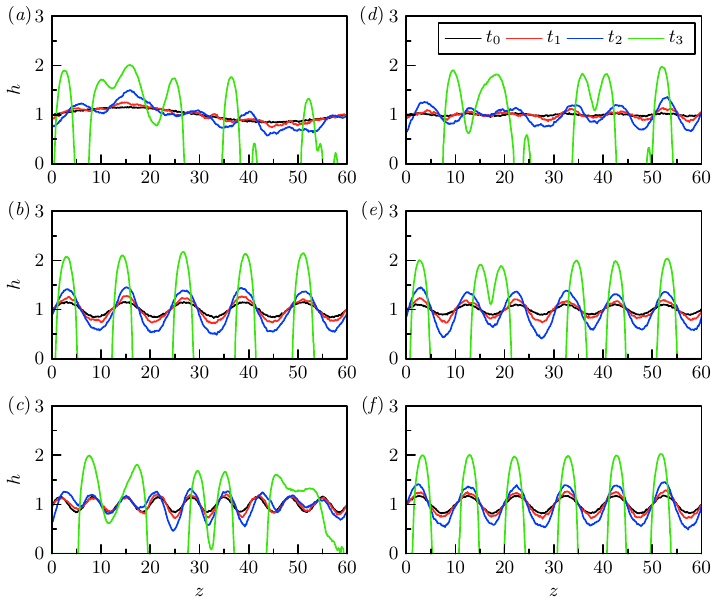}

	\caption{ 
		Interface profiles after rupture with various wavenumbers $k_0$ and amplitudes $A_0$ of external perturbations. 
		(\textit{a,b,c}) The initial amplitudes of the external perturbations are fixed, i.e. $A_0=0.15$.
		(\textit{d,e,f}) The wavenumbers of external perturbations are fixed , i.e. $k_0=\pi/5$.
		(\textit{a}) Results with $k_0=\pi/30$ at four instants: 0.09, 8.42, 16.83 and 28.52. 
		(\textit{b}) Results with $k_0=\pi/6$ at four instants: 0.09, 4.68, 8.88 and 17.77. 
		(\textit{c}) Results with $k_0=3\pi/10$ at four instants: 0.09, 6.55, 13.09 and 22.91.
		(\textit{d}) Results with $A_0= 0.025$ at four instants: 0.09, 8.88, 17.77 and 30.40. 
		(\textit{e}) Results with $A_0= 0.1$ at four instants: 0.09, 6.08, 12.16 and 20.11. 
		(\textit{f}) Results with $A_0= 0.175$ at four instants: 0.09, 4.21, 8.42 and 16.83.
	}
	\label{different wavenumber}
\end{figure}

Based on the results and discussions in \S\,\ref{sec4-1} and \S\,\ref{sec4-2}, the final interface profiles of the liquid nano-threads are determined by both thermal fluctuations and external perturbations. 
Figure\,\ref{different wavenumber} illustrates the influence of $k_0$ and $A_0$ of the external perturbations on the interface profiles extracted from the MD simulations.

In figures\,\ref{different wavenumber}(\textit{a--c}), the initial amplitude of the external perturbations is fixed ($A_0=0.15$) with various wavenumbers.
The uniform breakup only appears in the case with $k_0= \pi/6 $. 
According to equation\,\eqref{LE_dispersion}, the dimensionless growth rate in case\,5 is $0.146$, much larger than those with $k_0= \pi/30 $ ($\omega=0.064$) and $k_0= 3 \pi/10$ ($\omega=0.028$).
Starting from the same amplitude, the external perturbation with larger growth rate is better able to overwhelm the effects of the thermal fluctuations, leading to the results in the left panel of figure\,\ref{different wavenumber}. 

Figures\,\ref{different wavenumber}(\textit{d--f}) show the impact of different initial amplitudes with the same wavenumber ($k_0=\pi/5$), where the external perturbations have the same growth rate.
The maximum $A_0$ is found to enhance the hydrodynamic component of the instability, generating uniform droplets after the rupture, shown in figure\,\ref{different wavenumber}(\textit{f}), while the minimum $A_0$ is overwhelmed by the thermal fluctuations, leading to the non-uniform breakup in figure\,\ref{different wavenumber}(\textit{d}).
Interestingly, the rupture in figure\,\ref{different wavenumber}(\textit{e}) is ``quasi-uniform'' with only one droplet coalescence.
Note that this result is extracted from one selected realisation. 
Uniform breakup can also be found in other realisations of the case with $A_0= 0.1$, indicating a transition regime from the non-uniform breakup to the uniform breakup. 

\begin{figure}
	\centering
	\includegraphics{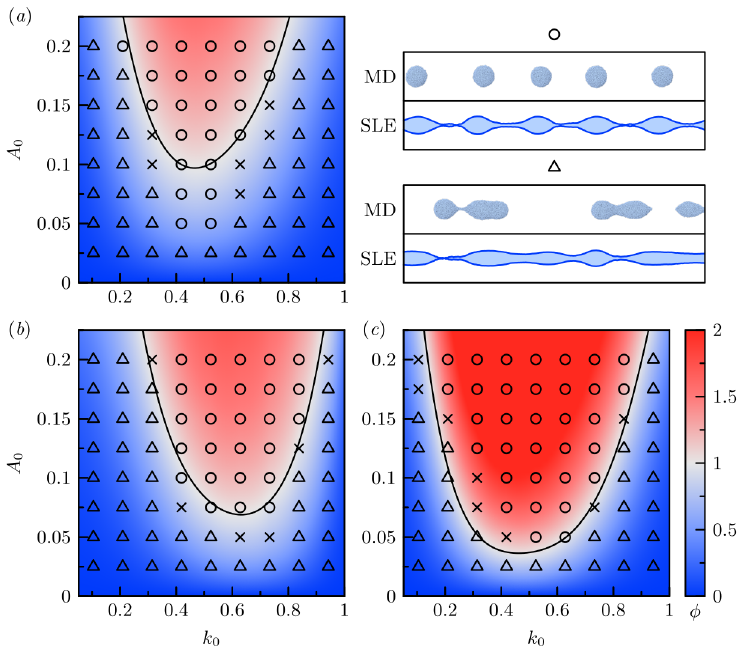}
	
	\caption{
		Regime maps at (\textit{a}) $\ohn=0.65$ and $\thn=0.07$, (\textit{b}) $\ohn=0.10$ and $\thn=0.07$, (\textit{c}) $\ohn=0.65$ and $\thn=0.04$. The regimes maps are depicted using contours of $\phi$ and symbols representing the numerical results obtained from (\textit{a}) MD and (\textit{b},\textit{c}) the SLE. Circles, triangles, and crosses denote the hydrodynamic, thermal-fluctuation and transition regimes, respectively.
	}
	\label{regime map}
\end{figure}

According to the simulation results in the preceding sections, two principal instability regimes can be summarised, providing a framework to describe different breakup patterns: (i) the ``hydrodynamic regime'', characterised by the generation of uniform droplets, and (ii) the ``thermal-fluctuation regime'', associated with non-uniform breakup.
To distinguish the regimes, a parameter $\phi$ is introduced to quantify the relative intensity of hydrodynamic component due to the external perturbations and thermal-fluctuation component, written as
\begin{equation}\label{criterion}
	\phi(t) = {\displaystyle \int_{0}^{\infty}\sqrt{\left\langle \left|\HL\right|^2 \right\rangle }\dd k} \bigg/ {\displaystyle \int_{0}^{\infty}\sqrt{\left\langle \left|\HF\right|^2 \right\rangle}\dd k}.
\end{equation}
Note that $\phi$ is time dependent. We set $\phi(t_r)=1$ as the boundary separating the hydrodynamic and thermal-fluctuation regimes. When $\phi(t_r)>1$, the external perturbations dominate the instability, while the thermal fluctuations exert more significant influence when $\phi(t_r)<1$.
For the fixed values of $\ohn$ and $\thn$, contours of $\phi$ are generated as a regime map based on $k_0$ and $A_0$, illustrated in figure\,\ref{regime map}.
Here, the distribution of $t_r$ in the regime map is fitted using a third-order polynomial based on the numerical results from the SLE.

Figure\,\ref{regime map}(\textit{a}) presents the regime map for the MD results ($\ohn = 0.65$ and $\thn = 0.07$). 
Besides the cases presented in figure\,\ref{different wavenumber}, more MD simulations with different values of $k_0$ and $A_0$ are performed to support the criterion of the regime map.
 Promisingly, the regime boundary (black solid line) from equation\,\eqref{criterion} generally matches the MD results represented by symbols (circles for the hydrodynamic regime and triangles for the thermal-fluctuation regime), except for the four circles at the bottom. 
The crosses suggest the transition regime, which emerge near the boundary, i.e. $\phi(t_r) \approx 1$.
This is consistent with the results of the case in figure\,\ref{different wavenumber}(\textit{e}).
Moreover, the bottom of the boundary indicates the optimum wavenumber ($k_0= 0.49$) for the hydrodynamic regime, closely matching the dominant mode predicted by the classical RP theory of equation\,\eqref{LE_dispersion}.

\begin{figure}
	\centering
	\includegraphics{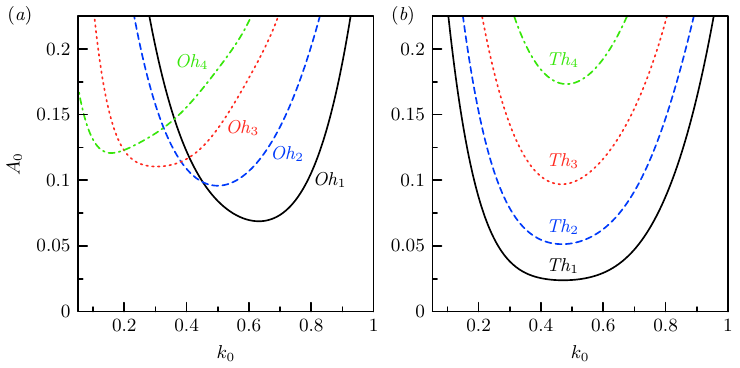}
	
	\caption{
		Regime boundaries at (\textit{a}) different $\ohn$ values ($\ohn_1=0.10$, $\ohn_2=0.50$, $\ohn_3=2.50$ and $\ohn_4=12.50$) with $\thn = 0.07$ and (\textit{b}) different $\thn$ values ($\thn_1=0.03$, $\thn_2=0.05$, $\thn_3=0.07$ and $\thn_4=0.09$) with $\ohn = 0.65$.
	}
	\label{regime boundary}
\end{figure}

Figures\,\ref{regime map}(\textit{b},\textit{c}) depict the influence of $\ohn$ and $\thn$ on the boundaries in the regime maps.
Since MD is not available for arbitrary values of $\ohn$ and $\thn$, numerical solutions of the SLE are employed to confirm the regime map across a broad range of $\thn$ and $\ohn$, specifically $\ohn = 0.10$ and $\thn = 0.07$ in figure\,\ref{regime map}(\textit{b}), $\ohn = 0.65$ and $\thn = 0.04$ in figure\,\ref{regime map}(\textit{c}).
Comparison between figures\,8(\textit{a},\textit{b}) reveals that reducing $\ohn$ results in a rightward shift of the regime boundary.
This trend is further presented in figure\,\ref{regime boundary}(\textit{a}) and can be explained by the classical RP theory, where the dominant wavenumber of the instability increases as $\ohn$ decreases.
Notably, the bottom points of the boundary in figure\,\ref{regime boundary}(\textit{a}) also exhibit a slight upward movement. The main reason is that $\ohn$ not only affects the hydrodynamic component but also modifies the intensity of thermal fluctuations, as shown in equation\,\eqref{Hrms}.
Examining figures\,\ref{regime map}(\textit{a},\textit{c}) and figure\,\ref{regime boundary}(\textit{b}), the regime boundary is observed to move downward as $\thn$ decreases, indicating that it becomes easier to enter the hydrodynamic regime with weaker thermal fluctuations.


\section{Conclusions}\label{sec5}
In this article, the SLE and MD are utilised to explore the influence of external perturbations and thermal fluctuations on the dynamics of liquid nano-threads.

Linear instability analysis is performed to derive a theoretical model for the spectra of thermal capillary waves, influenced by both thermal fluctuations and external perturbations.
This model, validated by MD simulations, reveals the instability mode of a spike from a specific external perturbation and a continuous curve due to thermal fluctuations, corresponding to the uniform and non-uniform ruptures, respectively.
An analytical model is then established for the two typical distributions of thermal capillary waves: bimodal distribution for uniform waves and Gaussian distribution for stochastic ones. 
Besides the formulation of thermal capillary waves, the evolution of 
perturbation growth, particularly in cases with uniform rupture, is also investigated.
The results of uniform rupture show that the perturbation grows exponentially at the linear stage, approximately following the classical linear theory proposed by \citet{eggers1994drop}, indicating the dominant roles of surface tension forces arising from the external perturbation with specific wavenumbers. 
However, The nonlinear evolution near rupture, determined jointly by surface tension forces and thermal fluctuations, is observed not be affected by the external perturbations.
Finally, two distinct regimes are defined to characterise the instability: (i) the hydrodynamic regime, marked by uniform droplets controlled by external perturbations, and (ii) the thermal-fluctuation regime, exhibiting a stochastic breakup pattern. A criterion is proposed to draw a regime map based on the perturbation amplitude ($A_0$) and wavenumber ($k_0$). The boundaries of these regimes, validated by MD and SLE simulations, are obtained including a transition area observed.

While this article provides new understanding of interfacial dynamics, it opens up several new avenues of enquiry.
One avenue involves deriving the Fokker-Planck equation of the SLE, a deterministic equation describing the probability density function of $\tilde{h}$. The utilisation of the Fokker-Planck equation holds the promise not only to fortify the mathematical underpinnings of the distribution function in \S\,\ref{sec4-1} but also to provide additional theoretical insights into the nonlinear dynamics of liquid nano-threads.
Another avenue is extending this study to a more practical fluid configuration, i.e. a liquid nano-jet. Despite the performed MD simulations for nano-jets \citep{moseler2000formation, choi2006molecular, kang2008thermal}, the introduction of external perturbations, widely employed at the macroscale \citep{yang2019manipulation, mu2023modulation}, remains unexplored for actively controlling the breakup of nano-jets.

\backsection[Acknowledgements]{
	Useful discussions with Dr. Kai Mu and Dr. Jingbang Liu are gratefully acknowledged.}

\backsection[Funding]{This work was supported by the National Natural Science Foundation of China (grant no. 12202437, 12027801, 12388101), the Youth Innovation Promotion Association CAS (grant no. 2018491), the Fundamental Research Funds for the Central Universities, the Opening fund of State Key Laboratory of Nonlinear Mechanics, the China Postdoctoral Science Foundation (grant no. 2022M723044) and the International Postdoctoral Fellowship (grant no. YJ20210177)}
\backsection[Declaration of interests]{
	The authors report no conflict of interest.}

\backsection[Author ORCIDs]{\\
	Chengxi Zhao \url{https://orcid.org/0000-0002-3041-0882}; \\
	Ting Si \url{https://orcid.org/0000-0001-9071-8646}.
}

\appendix

\section{Derivation of the Spectrum Function}\label{appc}

Equation\,\eqref{transformed LSLE} is a linear equation with time-invariant coefficients, so it satisfies the superposition principle of solutions, which supports us to decompose the full solution into two components
\begin{equation}\label{HL+HF}
	H=\HL+\HF.
\end{equation}
Considering the initial interface shape and assuming the initial velocity of the interface to be zero, the intial conditions are $\left.H\right|_{t=0}=H_\sub{i}$ and $\left. {\d H/\d t} \right|_{t=0}=0$.
$H_\sub{i}$ represents the spectrum of the initial interface profile $h(z,0)$.
The first term on the right side can be obtained by solving the homogeneous form of equation\,\eqref{transformed LSLE} with these initial conditions
\begin{equation}\label{HL}
	\HL=H_\sub{i}\exp\left( \frac{-a}{2}t \right) \left[ \cosh\left( \frac{b}{2}t \right) + \frac{a}{b}\sinh\left( \frac{b}{2}t \right) \right],
\end{equation}
where
\begin{equation*}
	a=3\ohn k^2, \quad b=\sqrt{\left(9\ohn^2 -2\right)k^4+2k^2}.
\end{equation*}
The second term on the right side of equation\,\eqref{HL+HF} could be calculated from the convolution of the excitation function (i.e. the inhomogeneous term) and the impulse response of equation\,\eqref{transformed LSLE}
\begin{equation}
	\HF=\sqrt{\frac{3\ohn}{2\pi}}\thn k^2 \int_{0}^{t} \xi(k, t-\tau) G(k, \tau) \dd\tau.
\end{equation}
The impulse response $G(k, t)$ could be obtained by solving the equation
\begin{equation}
	\frac{\d^2 G}{\d t^2}+ 3\ohn k^2 \frac{\d G}{\d t}+ \frac{k^4-k^2}{2} G=\delta(t),
\end{equation}
So we have
\begin{equation}
	G(k, t) = \frac{2}{b}\exp\left( \frac{-a}{2}t \right)\sinh\left( \frac{b}{2}t \right).
\end{equation}

As $H$ is a complex random variable with a zero mean, we should analyse it statistically, i.e. seek its root mean square from equation\,\eqref{HL+HF}
\begin{equation}
	\left|H\right|_\sub{rms} = \sqrt{\left\langle \left| \HL+\HF \right|^2 \right\rangle} = \sqrt{\left\langle \left|\HL\right|^2 \right\rangle + \left\langle \left|\HF\right|^2 \right\rangle},
\end{equation}
where the cross term is erased since $\HL$ and $\HF$ are orthogonal. Then, applying the same operation to equation\,\eqref{HL} readily yields
\begin{equation}
	\left\langle \left|\HL\right|^2 \right\rangle =\left|H_\sub{i}\right|^2\e^{-at} \left[ \cosh\left( \frac{b}{2}t \right) + \frac{a}{b}\sinh\left( \frac{b}{2}t \right) \right]^2.
\end{equation}

Given that $\xi(k,t)$ is an uncorrelated Gaussian white noise, we derive $\left\langle \left| \xi(k,t) \right|^2 \right\rangle = L$ which leads to
\begin{align}
	\left\langle \left|\HF\right|^2 \right\rangle &= \frac{3\ohn}{2\pi}\thn^2 k^4 \left\langle \left|\int_{0}^{t} \xi(k, t-\tau) G(k, \tau) \dd\tau\right|^2 \right\rangle  \notag \\
	&= \frac{3\ohn}{2\pi}\thn^2 k^4 \int_{0}^{t} \left\langle \left|\xi(k, t-\tau)\right|^2 \right\rangle G(k, \tau)^2 \dd\tau \notag \\
	&= \frac{3\ohn}{2\pi}\thn^2 k^4 L \int_{0}^{t} G(k, \tau)^2 \dd\tau \notag \\
	&= \frac{3L\ohn}{\pi}\thn^2 k^4\frac{a^2-b^2-a^2 \cosh (b t)-a c \sinh (b t)+b^2 \e^{a t}}{a b^2\left(a^2-b^2\right) \e^{a t}}.
\end{align}
Organising all the above results, we have the spectum function $\left|H\right|_\sub{rms}$ described in equations \eqref{Hrms} and \eqref{HLE and Hf}.

\section{Spectral leakage}\label{appd}

\begin{figure}
	\centering
	\includegraphics{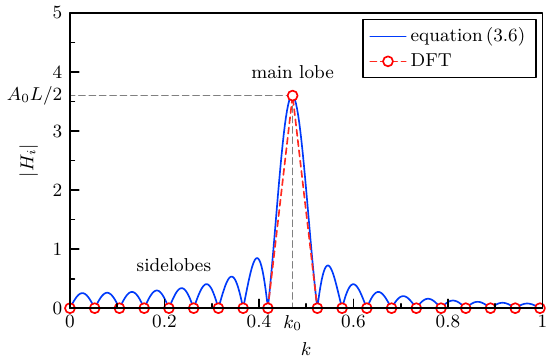} 
	
	\caption{
		Spectra of a sinusoidal wave $A_0\sin\left( k_0 z \right)$ with $z \in \left[0, L\right]$, where $A_0=0.06$, $k_0=3\pi/20$ and $L=120$. The results are obtained from equation\,\eqref{Hi} (blue solid line) and numerical DFT (red dashed line with circles).
	}
	\label{spectral leakage}
\end{figure}

The Fourier transform over a finite range $\left[0, L\right]$ introduces spectral leakage, leading to the prediction of numerous irrelevant modes (sidelobes) besides $k_0$ as depicted in figure\,\ref{spectral leakage} \citep{proakis1996digital}. However, in DFT, a finite signal is extended periodically, resulting in a discrete spectrum \citep{proakis1996digital} where the sidelobes cannot be captured.
To align with the outcomes obtained from the DFT, we eliminate these sidelobes and retain only the main lobe. The peak value of this main lobe is $\displaystyle \left| H_\sub{i} \right|_\sub{max} = \left| H_\sub{i} \right|_{k=k_0} = A_0L/2$, and its bandwidth is $4\pi/L$.

\section{Details of the MacCormack method}\label{appa}
Two differential operators, $\df$ and $\db$ are introduced to represent the forward and backward differences, respectively
\begin{equation}
		\df f = \frac{f_{i+1} - f_i}{z_{i+1} - z_i}, \quad
		\db f = \frac{f_i - f_{i-1}}{z_i - z_{i-1}}.
\end{equation}
$\bm{D}$ is discretised by the forward difference for the predictor step, written as
\begin{equation}
	\bm{D}\left(u^t_i,h^t_i\right) =
	\begin{pmatrix}
		D_1 \\[4pt]
		D_2  
	\end{pmatrix},
\end{equation}
where

\begin{empheq}[left={\empheqlbrace}]{alignat=1}
	D_1 = & -u^t_i \df u^t_i -\df p^t_i
	+ \frac{3\ohn}{\left(h^t_i\right)^2} \frac{\left(h^t_{i+1}\right)^2 \df u_i^t - \left(h^t_{i}\right)^2 \db u_i^t}{z_{i+1}-z_i} \notag \\ %
	& + \sqrt{\frac{6}{\pi}} \frac{\thn\sqrt{\ohn}} {\left(h^t_i\right)^2} \df \left(h^t_i N^t_i\right), \notag \\ %
	D_2 = & -u_i^t \df h^t_i - \frac{1}{2} h^t_i \df u^t_i. \notag %
\end{empheq}
The backward difference is applied for $\bar{\bm{D}}$
\begin{equation}
	\bar{\bm{D}}\left(\bar{u}^{t+1}_i, \bar{h}^{t+1}_i\right)=
	\begin{pmatrix}
		\bar{D}_1 \\[4pt]
		\bar{D}_2
	\end{pmatrix},
\end{equation}
where
\begin{empheq}[left={\empheqlbrace}]{alignat=1}
	\bar{D}_1 = & -\bar{u}^{t+1}_i \db \bar{u}^{t+1}_i -\db \bar{p}^{t+1}_i 
	+ \frac{3\ohn}{\left(\bar{h}^{t+1}_i\right)^2} \frac{\left(\bar{h}^{t+1}_{i}\right)^2 \df \bar{u}_i^{t+1} - \left(\bar{h}^{t+1}_{i-1}\right)^2 \db \bar{u}_i^{t+1}}{z_{i}-z_{i-1}} \notag \\ %
	& +\sqrt{\frac{6}{\pi}}\frac{\thn\sqrt{ \ohn }}{\left(\bar{h}^{t+1}_i\right)^2} \db\left(\bar{h}^{t+1}_i N^t_i\right), \notag \\ %
	\bar{D}_2 = & -\bar{u}_i^{t+1} \db \bar{h}^{t+1}_{i} - \frac{1}{2}\bar{h}^{t+1}_{i} \db \bar{u}^{t+1}_i. \notag %
\end{empheq}

\section{Spatially correlated noise model}\label{appb}
\begin{figure}
	\centering
	\includegraphics[width=0.6\textwidth]{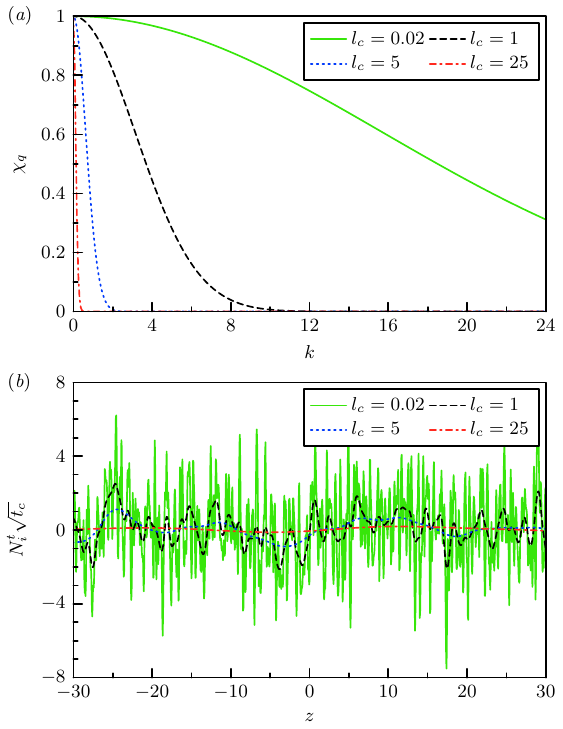} 
	
	\caption{
		(\textit{a}) Linear spectra of eigenvalues for several values of $l_\sub{c}$ from (\ref{eq_chi}). Here, the wavenumber, $k=2 \pi q/L$. 
		(\textit{b}) Spatially correlated noise with different $l_\sub{c}$.
	}
	\label{fig_app_noise}
\end{figure}

In this appendix, we introduce the spatially correlated noise model, first proposed by \citet{grun2006thin} for nanoscale bounded films, where an exponential correlation function is employed
\begin{equation}\label{eq_app_corre}
	F_\sub{c}(z,l_\sub{c})=
	\begin{cases}\displaystyle
		\frac{1}{X} \exp\left(-\frac{1}{2} \left[ \frac{L}{l_\sub{c}} \sin\left( \frac{\pi z}{L}\right)\right]^2 \right) & \text{for } l_\sub{c} > 0, \\
		\delta(z) & \text{for } l_\sub{c} = 0.     
	\end{cases}
\end{equation}
Here, $l_\sub{c}$ is the spatial correlation length, $L$ is the domain length, $X$ is such that $\int_0^L F_\sub{c}(z,l_\sub{c}) \dd z = 1$.
\citet{diez2016metallic} calculated the integral and found that $\chi_q$ could be expressed by the Bessel function
\begin{equation}
	\chi_q = \mathrm{I}_q(\alpha) \big/ \mathrm{I}_0(\alpha),
	\label{eq_chi}
\end{equation}
where
\begin{equation}
	\alpha = \left( \frac{L}{2 l_\sub{c}} \right)^2, \quad 
	k = \frac{2\pi q}{L}.
\end{equation}
Figure\,\ref{fig_app_noise}(\textit{a}) shows the eigenvalue spectra for several values of $l_\sub{c}$.
Note that for $l_\sub{c} \rightarrow 0$ (i.e., $\alpha \rightarrow \infty$), we have $\chi_q \rightarrow 1$ for all $q$, leading to the limiting case of the white (uncorrelated) noise.

The term $g_q$ corresponds to the set of orthonormal eigenfunctions according to
\begin{equation}\label{eq_fourier_domain_app2}
	g_\sub{q}(z)=
	\begin{cases}
		\displaystyle \sqrt{\frac{2}{L}} \cos\left( \frac{2\pi qz}{L} \right) & \text{for } q > 0, \\[8pt]
		\displaystyle \sqrt{\frac{1}{L}} & \text{for } q = 0, \\[8pt]
		\displaystyle \sqrt{\frac{2}{L}} \sin\left( \frac{2\pi qz}{L} \right) & \text{for } q < 0.
	\end{cases}
\end{equation}
Combining with the temporal correlated model proposed by \citet{zhao2020dynamics}, the final discretised expression of the noise term is 
\begin{equation}
	N^t_i= \frac{1}{\sqrt{t_\sub{c}}}\sum^{\frac{M+1}{2}}_{q=-\frac{M+1}{2}} \chi_q\, \mathcal{N}_q\, g_q(z) \,.
\end{equation}
Samples of $N^t_i$ at one time instant are illustrated in figure\,\ref{fig_app_noise}(\textit{b}) with different spatial correlation lengths.
Note that a larger $l_\sub{c}$ leads to smooth large-wavelength and small-amplitude noise.

	\section{Influence of the correlation lengths}\label{app_tc_lc}	
	
	In this appendix, we investigate the influence of the correlation length on the dynamics at both linear and nonlinear stages.
	
	\begin{figure}
		\centering
		\includegraphics[width=1.0\textwidth]{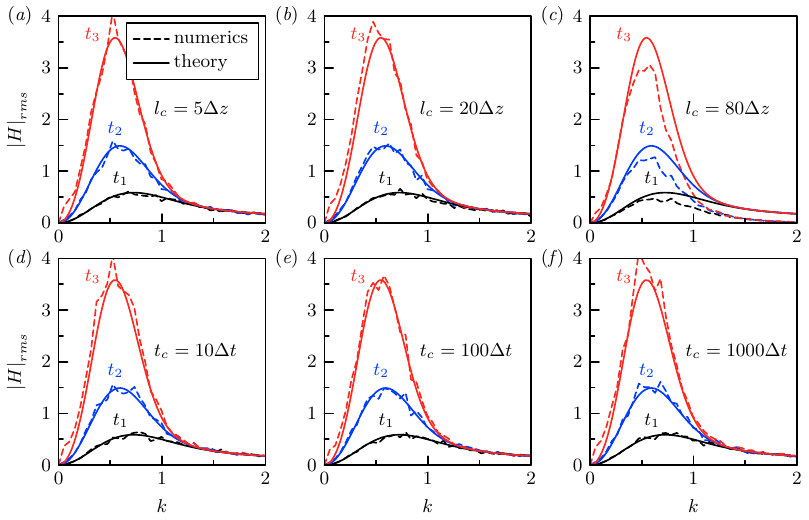} 
		
		\caption{
				Comparisons between theoretical (solid lines) and numerical (dashed lines with circles) spectra from the SLE at three instants: $t_1 = 6$ (black), $t_2 = 11 $ (blue) and $t_3 = 18 $ (red). (\textit{a}--\textit{c}) Influence of $l_c$ with $t_c=10 \rmDelta t$; (\textit{d}--\textit{f}) influence of $t_c$ with $l_c=5 \rmDelta z$. }
		\label{fig_app_spectra}
	\end{figure}
	
	For the linear instability, we conduct the SLE simulations by using different correlation lengths for a long thread with parameters from case\,1 ($L=120$, $\ohn=0.65$, $\thn=0.07$ and $A_0=0$).
	Using a similar approach employed in \S\,\ref{sec4-1}, 50 independent realisations are performed to gain statistics. 
	A discrete Fourier transform of the interface position is then applied to get the ensemble-averaged spectra.
	Figure\,\ref{fig_app_spectra} illustrates the influence of both the spatial correlation length $l_c$ and temporal one $t_c$.
	Comparing figures\,\ref{fig_app_spectra}(\textit{a}) and (\textit{b}), the spatial correlation length is not found to has a significant impact on spectra when $l_c \leq 20 \rmDelta z$. However, when $l_c = 80 \rmDelta z$, there is a notable reduction in the spectrum at high wavenumbers compared to the theoretical results, suggesting that a larger correlation length suppresses capillary waves driven by thermal fluctuations. Additionally, figures\,\ref{fig_app_spectra}(\textit{d}--\textit{f}) indicate that, for the time step ($\rmDelta t = 5 \times 10^{-6}$) used in this paper, $t_c$ within the range of $1000 \rmDelta t$ have no significant impact on the instability results.
	
\begin{figure}
	\centering
	\includegraphics[width=0.65\textwidth]{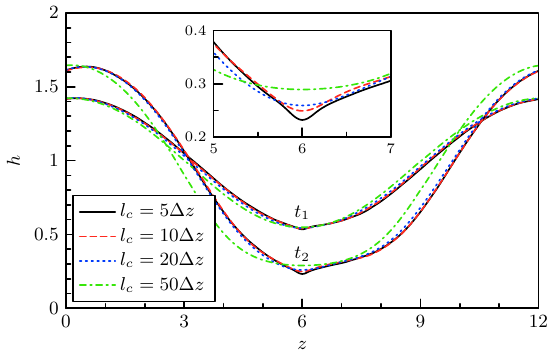} 
	
	\caption{
			Ensemble-averaged interface profiles at two instants ($t_1 = 1.5$ and $t_2 = 4.5$) influenced by $l_c$.
	}
	\label{fig_app_interface}
\end{figure}

	Furthermore, we examine the impacts of correlation lengths on the interface profiles, particularly at the nonlinear stage. Given that figure\,\ref{fig_app_spectra} demonstrates the minimal effects of $t_c$, only the influence of $l_c$ is explored here. To reduce computational costs, we consider the simulations of a short thread ($L=12$, $\ohn=0.65$, $\thn=0.07$ and $A_0=0.4$) with various spatial correlation lengths. Multiple SLE simulations (100 for each case) are then performed to get ensemble-averaged profiles.
	The results in figure\,\ref{fig_app_interface} indicate that when $l_c \leq 20 \rmDelta z$, the overall interface profiles are not significantly affected, aligning with the findings at the linear stage in figure\,\ref{fig_app_spectra}. However, the local interface morphology near $h_{min}$ is found to be affected by the spatial correlation lengths, indicating that $l_c$ in the numerical model represents the smallest spatial scale of thermal fluctuations.
	
	Therefore, we can conclude that variations in correlation length within a certain range do not significantly affect the computational results. They only influence the local behaviours of fluctuating hydrodynamics below the correlation length. In this study, we choose two relatively small correlation lengths, $l_\sub{c} = 5 \rmDelta z=0.15$ and $t_\sub{c} = 10 \rmDelta t = 5 \times 10^{-5}$, which approximately correspond to the molecular scale and a timescale of one femtosecond in MD simulations, respectively. These parameters essentially preserves the true physical characteristics of thermal fluctuations in physical space.

\bibliographystyle{jfm}
\bibliography{jfm}

\begin{thebibliography}{56}
\expandafter\ifx\csname natexlab\endcsname\relax\def\natexlab#1{#1}\fi
\def\au#1{#1} \def\ed#1{#1} \def\yr#1{#1}\def\at#1{#1}\def\jt#1{\textit{#1}}
  \def\bt#1{#1}\def\bvol#1{\textbf{#1}} \def\vol#1{#1} \def\pg#1{#1}
  \def\publ#1{#1}\def\arxiv#1{#1}\def\org#1{#1}\def\st#1{\textit{#1}}

\bibitem[Arienti {\em et~al.\/}(2011)Arienti, Pan, Li \&
  Karniadakis]{arienti2011many}
{\sc \au{Arienti, M.}, \au{Pan, W.}, \au{Li, X.} \& \au{Karniadakis, G.}}
  \yr{2011}  \at{Many-body dissipative particle dynamics simulation of
  liquid/vapor and liquid/solid interactions}.  \jt{J. Chem. Phys.}
  \bvol{134}~(20),  \pg{204114}.

\bibitem[Barker {\em et~al.\/}(2023)Barker, Bell \& Garcia]{Barker2023}
{\sc \au{Barker, B.}, \au{Bell, J.~B.} \& \au{Garcia, A.~L.}} \yr{2023}
  \at{Fluctuating hydrodynamics and the {Rayleigh–Plateau} instability}.
  \jt{Proc. Natl. Acad. Sci. U. S. A.}  \bvol{120}~(30),  \pg{e2306088120}.

\bibitem[Castrej{\'o}n-Pita {\em et~al.\/}(2015)Castrej{\'o}n-Pita,
  Castrej{\'o}n-Pita, Thete, Sambath, Hutchings, Hinch, Lister \&
  Basaran]{castrejon2015plethora}
{\sc \au{Castrej{\'o}n-Pita, J.}, \au{Castrej{\'o}n-Pita, A.~A.}, \au{Thete,
  S.~S.}, \au{Sambath, K.}, \au{Hutchings, I.~M.}, \au{Hinch, J.}, \au{Lister,
  J.~R.} \& \au{Basaran, O.~A.}} \yr{2015}  \at{Plethora of transitions during
  breakup of liquid filaments}.  \jt{Proc. Natl. Acad. Sci. U. S. A.}
  \bvol{112}~(15),  \pg{4582--4587}.

\bibitem[Chen \& Steen(1997)]{chen1997dynamics}
{\sc \au{Chen, Y.-J.} \& \au{Steen, P.~H.}} \yr{1997}  \at{Dynamics of inviscid
  capillary breakup: collapse and pinchoff of a film bridge}.  \jt{J. Fluid
  Mech.}  \bvol{341},  \pg{245--267}.

\bibitem[Choi {\em et~al.\/}(2006)Choi, Kim \& Kim]{choi2006molecular}
{\sc \au{Choi, Y.~S.}, \au{Kim, S.~J.} \& \au{Kim, M.-U.}} \yr{2006}
  \at{Molecular dynamics of unstable motions and capillary instability in
  liquid nanojets}.  \jt{Phys. Rev. E}  \bvol{73}~(1),  \pg{016309}.

\bibitem[Day {\em et~al.\/}(1998)Day, Hinch \& Lister]{day1998self}
{\sc \au{Day, R.~F.}, \au{Hinch, E.~J.} \& \au{Lister, J.~R.}} \yr{1998}
  \at{Self-similar capillary pinchoff of an inviscid fluid}.  \jt{Phys. Rev.
  Lett.}  \bvol{80}~(4),  \pg{704}.

\bibitem[Diez {\em et~al.\/}(2016)Diez, Gonz{\'a}lez \&
  Fern{\'a}ndez]{diez2016metallic}
{\sc \au{Diez, J.~A.}, \au{Gonz{\'a}lez, A.~G.} \& \au{Fern{\'a}ndez, R.}}
  \yr{2016}  \at{Metallic-thin-film instability with spatially correlated
  thermal noise}.  \jt{Phys. Rev. E}  \bvol{93}~(1),  \pg{013120}.

\bibitem[Eggers(1993)]{eggers1993universal}
{\sc \au{Eggers, J.}} \yr{1993}  \at{Universal pinching of {3D} axisymmetric
  free-surface flow}.  \jt{Phys. Rev. Lett.}  \bvol{71}~(21),  \pg{3458}.

\bibitem[Eggers(2002)]{eggers2002dynamics}
{\sc \au{Eggers, J.}} \yr{2002}  \at{Dynamics of liquid nanojets}.  \jt{Phys.
  Rev. Lett.}  \bvol{89}~(8),  \pg{084502}.

\bibitem[Eggers \& Dupont(1994)]{eggers1994drop}
{\sc \au{Eggers, J.} \& \au{Dupont, T.~F.}} \yr{1994}  \at{Drop formation in a
  one-dimensional approximation of the {Navier-Stokes} equation}.  \jt{J. Fluid
  Mech.}  \bvol{262},  \pg{205--221}.

\bibitem[Eggers \& Villermaux(2008)]{eggers2008physics}
{\sc \au{Eggers, J.} \& \au{Villermaux, E.}} \yr{2008}  \at{Physics of liquid
  jets}.  \jt{Rep. Prog. Phys.}  \bvol{71},  \pg{1--79}.

\bibitem[Fowlkes {\em et~al.\/}(2012)Fowlkes, Horton, Fuentes-Cabrera \&
  Rack]{fowlkes2012signatures}
{\sc \au{Fowlkes, J.}, \au{Horton, S.}, \au{Fuentes-Cabrera, M.} \& \au{Rack,
  P.~D.}} \yr{2012}  \at{Signatures of the {Rayleigh-Plateau} instability
  revealed by imposing synthetic perturbations on nanometer-sized liquid metals
  on substrates}.  \jt{Angew. Chem. Int. Ed.}  \bvol{51}~(35),
  \pg{8768--8772}.

\bibitem[Gopan \& Sathian(2014)]{gopan2014rayleigh}
{\sc \au{Gopan, N.} \& \au{Sathian, S.~P.}} \yr{2014}  \at{Rayleigh instability
  at small length scales}.  \jt{Phys. Rev. E}  \bvol{90}~(3),  \pg{033001}.

\bibitem[Green(1954)]{green1954markoff}
{\sc \au{Green, M.~S.}} \yr{1954}  \at{Markoff random processes and the
  statistical mechanics of time-dependent phenomena. {II}. {Irreversible}
  processes in fluids}.  \jt{J. Chem. Phys.}  \bvol{22}~(3),  \pg{398--413}.

\bibitem[Gr{\"u}n {\em et~al.\/}(2006)Gr{\"u}n, Mecke \&
  Rauscher]{grun2006thin}
{\sc \au{Gr{\"u}n, G.}, \au{Mecke, K.} \& \au{Rauscher, M.}} \yr{2006}
  \at{Thin-film flow influenced by thermal noise}.  \jt{J. Stat. Phys.}
  \bvol{122}~(6),  \pg{1261--1291}.

\bibitem[Hennequin {\em et~al.\/}(2006)Hennequin, Aarts, van~der Wiel, Wegdam,
  Eggers, Lekkerkerker \& Bonn]{hennequin2006drop}
{\sc \au{Hennequin, Y.}, \au{Aarts, D. G. A.~L.}, \au{van~der Wiel, J.~H.},
  \au{Wegdam, G.}, \au{Eggers, J.}, \au{Lekkerkerker, H. N.~W.} \& \au{Bonn,
  D.}} \yr{2006}  \at{Drop formation by thermal fluctuations at an ultralow
  surface tension}.  \jt{Phys. Rev. Lett.}  \bvol{97}~(24),  \pg{244502}.

\bibitem[Kadau {\em et~al.\/}(2007)Kadau, Rosenblatt, Barber, Germann, Huang,
  Carl{\`e}s \& Alder]{kadau2007importance}
{\sc \au{Kadau, K.}, \au{Rosenblatt, C.}, \au{Barber, J.~L.}, \au{Germann,
  T.~C.}, \au{Huang, Z.}, \au{Carl{\`e}s, P.} \& \au{Alder, B.~J.}} \yr{2007}
  \at{The importance of fluctuations in fluid mixing}.  \jt{Proc. Natl. Acad.
  Sci. U. S. A.}  \bvol{104}~(19),  \pg{7741--7745}.

\bibitem[Kang {\em et~al.\/}(2008)Kang, Landman \& Glezer]{kang2008thermal}
{\sc \au{Kang, W.}, \au{Landman, U.} \& \au{Glezer, A.}} \yr{2008}  \at{Thermal
  bending of nanojets: Molecular dynamics simulations of an asymmetrically
  heated nozzle}.  \jt{Appl. Phys. Lett.}  \bvol{93}~(12).

\bibitem[Kaufman {\em et~al.\/}(2012)Kaufman, Tao, Shabahang, Banaei, Deng,
  Liang, Johnson, Fink \& Abouraddy]{kaufman2012structured}
{\sc \au{Kaufman, J.~J.}, \au{Tao, G.}, \au{Shabahang, S.}, \au{Banaei, E.-H.},
  \au{Deng, D.~S.}, \au{Liang, X.}, \au{Johnson, S.~G.}, \au{Fink, Y.} \&
  \au{Abouraddy, A.~F.}} \yr{2012}  \at{Structured spheres generated by an
  in-fibre fluid instability}.  \jt{Nature}  \bvol{487}~(7408),  \pg{463--467}.

\bibitem[Kavokine {\em et~al.\/}(2021)Kavokine, Netz \&
  Bocquet]{kavokine2021fluids}
{\sc \au{Kavokine, N.}, \au{Netz, R.~R.} \& \au{Bocquet, L.}} \yr{2021}
  \at{Fluids at the nanoscale: From continuum to subcontinuum transport}.
  \jt{Annu. Rev. Fluid Mech.}  \bvol{53},  \pg{377--410}.

\bibitem[Kirkwood \& Buff(1949)]{kirkwood1949statistical}
{\sc \au{Kirkwood, J.~G.} \& \au{Buff, F.~P.}} \yr{1949}  \at{The statistical
  mechanical theory of surface tension}.  \jt{J. Chem. Phys.}  \bvol{17}~(3),
  \pg{338--343}.

\bibitem[Koplik \& Banavar(1993)]{koplik1993molecular}
{\sc \au{Koplik, J.} \& \au{Banavar, J.~R.}} \yr{1993}  \at{Molecular dynamics
  of interface rupture}.  \jt{Phys. Fluids A}  \bvol{5}~(3),  \pg{521--536}.

\bibitem[Kubo(1957)]{kubo1957statistical}
{\sc \au{Kubo, R.}} \yr{1957}  \at{Statistical-mechanical theory of
  irreversible processes. {I}. {General} theory and simple applications to
  magnetic and conduction problems}.  \jt{J. Phys. Soc. Jpn.}  \bvol{12}~(6),
  \pg{570--586}.

\bibitem[Lagarde {\em et~al.\/}(2018)Lagarde, Josserand \&
  Proti{\`e}re]{lagarde2018oscillating}
{\sc \au{Lagarde, A.}, \au{Josserand, C.} \& \au{Proti{\`e}re, S.}} \yr{2018}
  \at{Oscillating path between self-similarities in liquid pinch-off}.
  \jt{Proc. Natl. Acad. Sci. U. S. A.}  \bvol{115}~(49),  \pg{12371--12376}.

\bibitem[Landau {\em et~al.\/}(1987)Landau, Lifshitz \&
  Pitaevskij]{landau1987statistical}
{\sc \au{Landau, L.~D.}, \au{Lifshitz, E.~M.} \& \au{Pitaevskij, L.~P.}}
  \yr{1987} {\em Statistical physics, part 2: theory of the condensed state\/}.
   \publ{Pergamon, New York}.

\bibitem[Li {\em et~al.\/}(2015)Li, Zhao, Mao, Wu \& Xu]{li2015focused}
{\sc \au{Li, C.}, \au{Zhao, L.}, \au{Mao, Y.}, \au{Wu, W.} \& \au{Xu, J.}}
  \yr{2015}  \at{Focused-ion-beam induced {Rayleigh-Plateau} instability for
  diversiform suspended nanostructure fabrication}.  \jt{Sci. Rep.}
  \bvol{5}~(1),  \pg{1--8}.

\bibitem[Liu {\em et~al.\/}(2023)Liu, Zhao, Lockerby \&
  Sprittles]{liu2023thermal}
{\sc \au{Liu, J.}, \au{Zhao, C.}, \au{Lockerby, D.~A.} \& \au{Sprittles,
  J.~E.}} \yr{2023}  \at{Thermal capillary waves on bounded nanoscale thin
  films}.  \jt{Phys. Rev. E}  \bvol{107}~(1).

\bibitem[Lord~Rayleigh(1878)]{rayleigh1878instability}
{\sc \au{Lord~Rayleigh, F. R.~S.}} \yr{1878}  \at{On the instability of jets}.
  \jt{Proc. Lond. Math. Soc.}  \bvol{10},  \pg{4--13}.

\bibitem[MacCormack(2003)]{maccormack2003effect}
{\sc \au{MacCormack, R.}} \yr{2003}  \at{The effect of viscosity in
  hypervelocity impact cratering}.  \jt{J. Spacecr. Rockets}  \bvol{40},
  \pg{757--763}.

\bibitem[Min \& Wong(2006)]{min2006rayleigh}
{\sc \au{Min, D.} \& \au{Wong, H.}} \yr{2006}  \at{Rayleigh's instability of
  {Lennard-Jones} liquid nanothreads simulated by molecular dynamics}.
  \jt{Phys. Fluids}  \bvol{18}~(2),  \pg{024103}.

\bibitem[Mo {\em et~al.\/}(2016)Mo, Qin, Zhao \& Yang]{mo2016application}
{\sc \au{Mo, C.}, \au{Qin, L.}, \au{Zhao, F.} \& \au{Yang, L.}} \yr{2016}
  \at{Application of the dissipative particle dynamics method to the
  instability problem of a liquid thread}.  \jt{Phys. Rev. E}  \bvol{94}~(6),
  \pg{063113}.

\bibitem[Molinero \& Moore(2009)]{molinero2009water}
{\sc \au{Molinero, V.} \& \au{Moore, E.~B.}} \yr{2009}  \at{Water modeled as an
  intermediate element between carbon and silicon}.  \jt{J. Phys. Chem. B}
  \bvol{113}~(13),  \pg{4008--4016}.

\bibitem[Moseler \& Landman(2000)]{moseler2000formation}
{\sc \au{Moseler, M.} \& \au{Landman, U.}} \yr{2000}  \at{Formation, stability,
  and breakup of nanojets}.  \jt{Science}  \bvol{289}~(5482),  \pg{1165--1169}.

\bibitem[Mu {\em et~al.\/}(2023)Mu, Qiao, Ding \& Si]{mu2023modulation}
{\sc \au{Mu, K.}, \au{Qiao, R.}, \au{Ding, H.} \& \au{Si, T.}} \yr{2023}
  \at{Modulation of coaxial cone-jet instability in active co-flow focusing}.
  \jt{J. Fluid Mech.}  \bvol{977},  \pg{A14}.

\bibitem[Papageorgiou(1995)]{papageorgiou1995breakup}
{\sc \au{Papageorgiou, D.~T.}} \yr{1995}  \at{On the breakup of viscous liquid
  threads}.  \jt{Phys. Fluids}  \bvol{7}~(7),  \pg{1529--1544}.

\bibitem[Papoulis \& Pillai(2002)]{edition2002probability}
{\sc \au{Papoulis, A.} \& \au{Pillai, S.~U.}} \yr{2002} {\em Probability,
  random variables, and stochastic processes\/}.  \publ{McGraw-Hill Europe: New
  York, NY, USA}.

\bibitem[Perumanath {\em et~al.\/}(2019)Perumanath, Borg, Chubynsky, Sprittles
  \& Reese]{perumanath2019droplet}
{\sc \au{Perumanath, S.}, \au{Borg, M.~K.}, \au{Chubynsky, M.~V.},
  \au{Sprittles, J.~E.} \& \au{Reese, J.~M.}} \yr{2019}  \at{Droplet
  coalescence is initiated by thermal motion}.  \jt{Phys. Rev. Lett.}
  \bvol{122}~(10),  \pg{104501}.

\bibitem[Petit {\em et~al.\/}(2012)Petit, Rivi{\`e}re, Kellay \&
  Delville]{petit2012break}
{\sc \au{Petit, J.}, \au{Rivi{\`e}re, D.}, \au{Kellay, H.} \& \au{Delville,
  J.-P.}} \yr{2012}  \at{Break-up dynamics of fluctuating liquid threads}.
  \jt{Proc. Natl. Acad. Sci. U. S. A.}  \bvol{109}~(45),  \pg{18327--18331}.

\bibitem[Plateau(1873)]{plateau1873statique}
{\sc \au{Plateau, J. A.~F.}} \yr{1873} {\em Statique exp{\'e}rimentale et
  th{\'e}orique des liquides soumis aux seules forces mol{\'e}culaires\/}.
  \publ{Gauthier-Villars}.

\bibitem[Proakis \& Manolakis(1996)]{proakis1996digital}
{\sc \au{Proakis, J.~G.} \& \au{Manolakis, D.~G.}} \yr{1996} {\em Digital
  signal processing: principles, algorithms, and applications\/}, 3rd edn.
  \publ{Prentice-Hall, Inc.}

\bibitem[Rice(2007)]{rice2007mathematical}
{\sc \au{Rice, J.~A.}} \yr{2007} {\em Mathematical statistics and data
  analysis\/}, 3rd edn.  \publ{Duxbury}.

\bibitem[Shah {\em et~al.\/}(2019)Shah, van Steijn, Kleijn \&
  Kreutzer]{shah2019thermal}
{\sc \au{Shah, M.~S.}, \au{van Steijn, V.}, \au{Kleijn, C.~R.} \& \au{Kreutzer,
  M.~T.}} \yr{2019}  \at{Thermal fluctuations in capillary thinning of thin
  liquid films}.  \jt{J. Fluid Mech.}  \bvol{876},  \pg{1090--1107}.

\bibitem[Shi {\em et~al.\/}(1994)Shi, Brenner \& Nagel]{shi1994cascade}
{\sc \au{Shi, X.~D.}, \au{Brenner, M.~P.} \& \au{Nagel, S.~R.}} \yr{1994}
  \at{A cascade of structure in a drop falling from a faucet}.  \jt{Science}
  \bvol{265}~(5169),  \pg{219--222}.

\bibitem[Thompson {\em et~al.\/}(2022)Thompson, Aktulga, Berger, Bolintineanu,
  Brown, Crozier, in't Veld, Kohlmeyer, Moore, Nguyen, Shan, Stevens,
  Tranchida, Trott \& Plimpton]{thompson2022lammps}
{\sc \au{Thompson, A.~P.}, \au{Aktulga, H.~M.}, \au{Berger, R.},
  \au{Bolintineanu, D.~S.}, \au{Brown, W.~M.}, \au{Crozier, P.~S.}, \au{in't
  Veld, P.~J.}, \au{Kohlmeyer, A.}, \au{Moore, S.~G.}, \au{Nguyen, T.~D.},
  \au{Shan, R.}, \au{Stevens, M.~J.}, \au{Tranchida, J.}, \au{Trott, C.} \&
  \au{Plimpton, S.~J.}} \yr{2022}  \at{{LAMMPS}--a flexible simulation tool for
  particle-based materials modeling at the atomic, meso, and continuum scales}.
   \jt{Comput. Phys. Commun.}  \bvol{271},  \pg{108171}.

\bibitem[Tiwari {\em et~al.\/}(2008)Tiwari, Reddy, Mukhopadhyay \&
  Abraham]{tiwari2008simulations}
{\sc \au{Tiwari, A.}, \au{Reddy, H.}, \au{Mukhopadhyay, S.} \& \au{Abraham,
  J.}} \yr{2008}  \at{Simulations of liquid nanocylinder breakup with
  dissipative particle dynamics}.  \jt{Phys. Rev. E}  \bvol{78}~(1),
  \pg{016305}.

\bibitem[Xue {\em et~al.\/}(2018)Xue, Sbragaglia, Biferale \&
  Toschi]{xue2018effects}
{\sc \au{Xue, X.}, \au{Sbragaglia, M.}, \au{Biferale, L.} \& \au{Toschi, F.}}
  \yr{2018}  \at{Effects of thermal fluctuations in the fragmentation of a
  nanoligament}.  \jt{Phys. Rev. E}  \bvol{98}~(1),  \pg{012802}.

\bibitem[Yang {\em et~al.\/}(2019)Yang, Qiao, Mu, Zhu, Xu \&
  Si]{yang2019manipulation}
{\sc \au{Yang, C.}, \au{Qiao, R.}, \au{Mu, K.}, \au{Zhu, Z.}, \au{Xu, R.~X.} \&
  \au{Si, T.}} \yr{2019}  \at{Manipulation of jet breakup length and droplet
  size in axisymmetric flow focusing upon actuation}.  \jt{Phys. Fluids}
  \bvol{31}~(9),  \pg{091702}.

\bibitem[Zhang {\em et~al.\/}(2016)Zhang, He, Li, Xu \& Li]{zhang2016micro}
{\sc \au{Zhang, B.}, \au{He, J.}, \au{Li, X.}, \au{Xu, F.} \& \au{Li, D.}}
  \yr{2016}  \at{Micro/nanoscale electrohydrodynamic printing: From {2D} to
  {3D}}.  \jt{Nanoscale}  \bvol{8}~(34),  \pg{15376--15388}.

\bibitem[Zhang {\em et~al.\/}(2019)Zhang, Sprittles \&
  Lockerby]{zhang2019molecular}
{\sc \au{Zhang, Y.}, \au{Sprittles, J.~E.} \& \au{Lockerby, D.~A.}} \yr{2019}
  \at{Molecular simulation of thin liquid films: {Thermal} fluctuations and
  instability}.  \jt{Phys. Rev. E}  \bvol{100}~(2),  \pg{023108}.

\bibitem[Zhao {\em et~al.\/}(2022)Zhao, Liu, Lockerby \&
  Sprittles]{zhao2022fluctuation}
{\sc \au{Zhao, C.}, \au{Liu, J.}, \au{Lockerby, D.~A.} \& \au{Sprittles,
  J.~E.}} \yr{2022}  \at{Fluctuation-driven dynamics in nanoscale thin-film
  flows: Physical insights from numerical investigations}.  \jt{Physical Review
  Fluids}  \bvol{7}~(2),  \pg{024203}.

\bibitem[Zhao {\em et~al.\/}(2020{\natexlab{{\em a\/}}})Zhao, Lockerby \&
  Sprittles]{zhao2020dynamics}
{\sc \au{Zhao, C.}, \au{Lockerby, D.~A.} \& \au{Sprittles, J.~E.}}
  \yr{2020{\natexlab{{\em a\/}}}}  \at{Dynamics of liquid nanothreads:
  {Fluctuation-driven} instability and rupture}.  \jt{Phys. Rev. Fluids}
  \bvol{5}~(4),  \pg{044201}.

\bibitem[Zhao {\em et~al.\/}(2019)Zhao, Sprittles \&
  Lockerby]{zhao2019revisiting}
{\sc \au{Zhao, C.}, \au{Sprittles, J.~E.} \& \au{Lockerby, D.~A.}} \yr{2019}
  \at{Revisiting the {Rayleigh-Plateau} instability for the nanoscale}.  \jt{J.
  Fluid Mech.}  \bvol{861},  \pg{R3}.

\bibitem[Zhao {\em et~al.\/}(2023)Zhao, Zhang \& Si]{zhao2023fluctuation}
{\sc \au{Zhao, C.}, \au{Zhang, Z.} \& \au{Si, T.}} \yr{2023}
  \at{Fluctuation-driven instability of nanoscale liquid films on chemically
  heterogeneous substrates}.  \jt{Phys. Fluids}  \bvol{35}~(7),  \pg{072016}.

\bibitem[Zhao {\em et~al.\/}(2021{\natexlab{{\em a\/}}})Zhao, Zhao, Si \&
  Chen]{zhao2021influence}
{\sc \au{Zhao, C.}, \au{Zhao, J.}, \au{Si, T.} \& \au{Chen, S.}}
  \yr{2021{\natexlab{{\em a\/}}}}  \at{Influence of thermal fluctuations on
  nanoscale free-surface flows: {A} many-body dissipative particle dynamics
  study}.  \jt{Phys. Fluids}  \bvol{33}~(11),  \pg{112004}.

\bibitem[Zhao {\em et~al.\/}(2020{\natexlab{{\em b\/}}})Zhao, Zhou, Zhang,
  Chen, Liu \& Wang]{zhao2020rupture}
{\sc \au{Zhao, J.}, \au{Zhou, N.}, \au{Zhang, K.}, \au{Chen, S.}, \au{Liu, Y.}
  \& \au{Wang, Y.}} \yr{2020{\natexlab{{\em b\/}}}}  \at{Rupture process of
  liquid bridges: {The} effects of thermal fluctuations}.  \jt{Phys. Rev. E}
  \bvol{102}~(2),  \pg{023116}.

\bibitem[Zhao {\em et~al.\/}(2021{\natexlab{{\em b\/}}})Zhao, Wan, Chen, Chao
  \& Xu]{zhao2021uniform}
{\sc \au{Zhao, Y.}, \au{Wan, D.}, \au{Chen, X.}, \au{Chao, X.} \& \au{Xu, H.}}
  \yr{2021{\natexlab{{\em b\/}}}}  \at{Uniform breaking of liquid-jets by
  modulated laser heating}.  \jt{Phys. Fluids}  \bvol{33}~(4),  \pg{044115}.

\end{thebibliography}

\end{document}